\documentclass[12pt]{article}

\usepackage{multicol,multirow}
\usepackage{amsmath,amssymb,cite}
\usepackage{graphicx}

\def\lsi{\raise0.3ex\hbox{$<$\kern-0.75em\raise-1.1ex\hbox{$\sim$}}}
\def\gsi{\raise0.3ex\hbox{$>$\kern-0.75em\raise-1.1ex\hbox{$\sim$}}}
\newcommand{\lsim}{\mathop{\lsi}}
\newcommand{\gsim}{\mathop{\gsi}}

\makeatletter
\@addtoreset{equation}{section}
\makeatother

\begin{document}

\begin{flushright}
HU-EP-05/68, CPT-2005/P.074 \\
BI-TP 2005/45,
DESY 05-217 \\
SFB/CPP-05-72 \\
\end{flushright}

\begin{center}

{\Large\bf Exploring Topology Conserving} \\
\vspace*{3mm}
{\Large\bf Gauge Actions for Lattice QCD} \\

\vspace*{7mm}

%AUTHORS
W. Bietenholz$^{1}$, K. Jansen$^{2}$, K.-I. Nagai$^{2}$, \\
S. Necco$^{3,4}$, L. Scorzato$^{1}$ and S. Shcheredin$^{5}$ \\

\vspace*{5mm}

{\small
$^{1}$ Institut f\"{u}r Physik, Humboldt Universit\"{a}t zu Berlin \\ 
Newtonstr. 15, D-12489 Berlin, Germany \\
\vspace*{2mm}
$^{2}$  NIC, DESY, Platanenallee 6 \\
 D-15738 Zeuthen, Germany \\
\vspace*{2mm}
$^{3}$ Centre de Physique Th\'{e}orique, Luminy, Case 907 \\ 
F-13288, Marseille Cedex 9, France \\
\vspace*{2mm}
$^{4}$ IFIC - Instituto de Fisica Corpuscular \\
Edificio Institutos de Investigacion \\
Apartado de Correos 22085 \\
E-46071 Valencia, Spain \\
\vspace*{2mm}
$^{5}$ Fakult\"{a}t f\"{u}r Physik, Universit\"{a}t Bielefeld \\
D-33615 Bielefeld, Germany 
}
\end{center}

\vspace*{5mm}

We explore gauge actions for lattice QCD, which are constructed
such that the occurrence of small plaquette values is strongly 
suppressed. By choosing strong bare gauge couplings
we arrive at values for the physical lattice spacings of 
${\cal O}(0.1 ~{\rm fm})$.
Such gauge actions tend to confine the Monte Carlo history to a
single topological sector.
This topological stability facilitates the collection of a large
set of configurations in a specific sector, which
is profitable for numerical studies in the $\epsilon$-regime. 
The suppression of small plaquette values is also expected to
be favourable for simulations with dynamical quarks.
We use a local Hybrid Monte Carlo algorithm to simulate such actions, 
and we present numerical results for the static potential, 
the physical scale, the topological stability
and the kernel condition number of the overlap Dirac operator.
In addition we discuss the question of reflection
positivity for a class of such gauge actions.

\newpage

\section{Introduction}

The gauge action of QCD can be lattice discretised
in many ways. One requires the naive continuum limit
for all lattice gauge actions to coincide, and one usually
assumes that then they fall into the same universality 
class. The formulation of the lattice gauge actions
is considered less problematic than the fermionic part of QCD,
in particular due to the absence of additive mass renormalisation.
%Thanks to asymptotic freedom, the glueball mass can be driven
%arbitrarily close to criticality without conceptual obstacles.
%A related property is that even 
Even for the simplest plaquette action,
known as the {\em Wilson gauge action} \cite{Wilson}, 
\footnote{$U_{P}$ denotes the plaquette variables in a lattice
gauge configurations given by the link variables $U_{x, \mu} \in SU(3)$.
The sum in eq.\ (\ref{Wilact}) runs over all plaquettes, see
e.g.\ Ref.\ \cite{MoMu}.}
\begin{equation}  \label{Wilact}
S_{\rm W}[U] = \beta \sum_{P} S_{P}(U_{P}) \ , \quad
S_{P}(U_{P}) = 1 - \frac{1}{3} {\rm Re~Tr} U_{P} \ ,
\end{equation}
the lattice artifacts in the scaling
behaviour only appear in ${\cal O}(a^{2})$, where $a$ is the 
lattice spacing. This is in contrast to the Wilson fermion,
which suffers from ${\cal O}(a)$ artifacts, as well as
additive mass renormalisation, so that criticality can only
be approached with a tedious fine tuning.

Nevertheless there are a number of suggestions for improved
lattice gauge actions. In general the goal is to further suppress the scaling 
artifacts by including more extended closed loops
in the discrete formulation of the field strength tensor.
This is achieved for instance by the tree level improved Symanzik gauge 
action \cite{PW}, the on-shell improved
L\"{u}scher-Weisz action \cite{LW}, and by several renormalisation group 
improved actions \cite{Iwa,DBW2,FPA}.

In the current work we also consider non-standard gauge actions
for lattice QCD, but we focus on another property, which we want to 
improve. Our intention is to suppress as far as possible the
occurrence of small plaquette values. Of course, this suppression
prevents the gauge field from fluctuating as it would for the
Wilson gauge action, hence we have to use much stronger bare gauge 
couplings to arrive at a comparable lattice spacing. For practical
simulations a lattice spacing of ${\cal O}(0.1 ~{\rm fm})$ is required;
this enables the use of a sensible
physical volume with lattices of a tractable size.
For quenched simulations with the Wilson gauge action such a 
lattice spacing is obtained for
$\beta = 6/g_{0}^{2} \approx 6$. We will see that a lattice spacing
of the same magnitude can still be obtained for the actions that
drastically suppress small plaquette values,
if we  drive $\beta$ down to values around and even below 1.

This observation means that these gauge actions can in fact be used in
simulations. Now we would like to explain what virtues we expect 
from it. % These motivations follow two distinct lines.

\section{Motivations for the suppression of small plaquette values}

Small plaquette values $U_{P}$ are expected 
to be linked to small eigenvalues of the Dirac operator, which is
most relevant in view of {\em dynamical simulations}.
On the practical side, the suppression of small values for $U_{P}$ should 
speed up dynamical Hybrid Monte Carlo (HMC)
simulations, avoid problems with the numerical integrator 
and ease the access to light pions. This would mean a
further improvement in this direction, in addition to 
recent algorithmic developments \cite{MLdyn}.
%In view of {\em dynamical simulations}, small plaquette values cause a number
%of problems: on the practical side, the generation of configurations 
%slows down, since small eigenvalues are involved and
%the evaluation of the fermion determinant takes 
%a long time. On the conceptual side it makes it hard to simulate
%light pions. In most simulations using Wilson fermions, the 
%pion mass is far above its physical value, which obstructs the numerical
%study of low energy QCD. Hence we hope that a suppression of small 
%plaquette values speeds up the simulations with dynamical quarks
%and provides access to lighter pions. 
However, these properties are not tested in the current work; we leave
them for further investigation.\\

We do investigate, however, another virtue, which we want to explain
now. There is a notorious lack of analytical tools to handle QCD
at low energy, so one often switches to the use of an effective
Lagrangian instead. In particular, {\em chiral perturbation theory}
($\chi$PT) is an effective theory in terms of the  Nambu-Goldstone 
boson fields of chiral symmetry breaking \cite{XPT}.
This approach is very powerful, and it still works if one
includes the small explicit chiral symmetry breaking due to the finite masses
of the light quarks. Then one deals with light pseudo Nambu-Goldstone
bosons, which are identified with the light mesons. 
%One performs a low energy expansion in terms of the momenta and masses
%of these mesons. This effective description can be applied in different
%fields of physics, and it leads to many important
%predictions. 
However, by its nature $\chi$PT cannot capture all the properties
of QCD as the fundamental theory.
%of the underlying, fundamental theory --- which is QCD in this case.
To supplement the missing information,
in particular the values of the Low Energy Constants
in the chiral Lagrangian, one has to relate $\chi$PT
to QCD results, which one tries to obtain from lattice simulations,
%Such an additional input is the evaluation of the coefficients in the
%terms of the effective chiral Lagrangian: these terms are classified 
%according to certain counting rules, but their coefficients 
%--- the Low Energy Constants --- appear
%in chiral perturbation theory as free parameters. They do play, however,
%an important r\^{o}le in physics, so they should be evaluated based on
%QCD, 
so we need simulations with light mesons.

In the recent years, a lattice fermion formulation became available,
which overcomes the conceptual problems related to the chiral symmetry.
This is the case if a lattice Dirac operator $D$ obeys the Ginsparg-Wilson 
(GW) relation \cite{GW,Has}. 
In its simplest form (and in lattice units, $a=1$) the GW relation reads
\begin{equation}
D  \gamma_{5} +  \gamma_{5} D = \frac{1}{\mu} D  \gamma_{5} D \ ,
\qquad \mu \gsim 1 \ .
\end{equation}
The corresponding GW fermions have a lattice modified, but exact chiral
symmetry at arbitrary lattice spacing, with the full number of 
generators \cite{ML}. This property rules out additive mass renormalisation
(along with ${\cal O}(a)$ scaling artifacts). Hence a small (bare)
quark mass $m_{q}$ does indeed imply a small pion mass, 
$m_{\pi}^{2} \propto m_{q}$. 

In practice, there are still difficulties left when we want
to approach the physical pion mass. To illuminate this point, we look
at the Neuberger overlap operator \cite{Neu1}, which is an often applied,
explicit solution to the GW relation. At zero fermion mass it takes the form
\begin{equation}  \label{overlap}
D_{\rm ov}^{(0)} =\mu \Big[ 1 + \gamma_{5} Q / \sqrt{Q^{2}} \Big] 
 \ , \quad
Q = \gamma_{5} (D_{\rm W} - \mu) \ ,
\end{equation}
where $D_{\rm W}$ is the Wilson Dirac operator. Its property
$D_{\rm W}^{\dagger} = \gamma_{5} D_{\rm W} \gamma_{5}$ implies
that $Q$ is Hermitian. A quark mass is added to the overlap operator
as follows,
\begin{equation} 
D_{\rm ov}(m_q ) =
\Big( 1 - \frac{m_{q}}{2 \mu} \Big) D_{\rm ov}^{(0)} + m_{q} \ .
\end{equation}
Simulations with this operator are computationally
expensive, so that the corresponding QCD studies are 
essentially restricted to
the quenched approximation up to now. The trouble-maker is the
inverse square root, which has to be approximated
by polynomials of a high degree. The effort to handle
$D_{\rm ov}$ is roughly
proportional to this degree, which grows --- for a fixed
accuracy of $D_{\rm ov}$ --- like the square root of the condition number
of $Q^{2}$. (One usually projects out the lowest few modes
and treats them separately to lower this condition number, but
this takes time again.) The lowest eigenvalues of $Q^{2}$ are again
expected to be linked to the occurrence of low plaquette values.
We are going to demonstrate an improvement also in this respect
for the gauge actions that we are going to consider.

Another obvious problem at small pion mass are the finite size effects.
In a box of side length $L$ they depend on the product $m_{\pi}L$,
which should be large to keep the finite size effects small, i.e.\
to be close to the physically realistic situation.
Then we are in the {\em $p$-regime}, where the $p$-expansion
of $\chi$PT \cite{p-reg} applies.
For instance, for $a = 0.1 ~{\rm fm}$ 
a pion mass of $250 ~{\rm MeV}$ corresponds to a
correlation length $\xi = 1/m_{\pi}$ of about 8 lattice spacings,
and to dwarf the finite size effects $L$ should be much larger.
%, say about $L \geq 4 \ \xi \, $. 
Lattices of such an extent are very 
expensive with GW fermions, even quenched, still without reaching 
physical pion masses.

As a way out one may consider the opposite situation,
\begin{equation}
\xi \leq L \ ,
\end{equation}
where we are confronted with strong finite size effects.
This setting is called the {\em $\epsilon$-regime}; it allows for the use
of a variant of chiral perturbation theory known as the 
$\epsilon$-expansion \cite{eps-reg}.
%\footnote{Further analytical work with the $\epsilon$ expansion
%can be found in Refs.\ \cite{eps-reg2,LeuSmi}.} 
It describes the finite size effects analytically, and its
formulae can be related to the numerically measured finite size effects.
The interesting point is that these formulae involve the Low Energy 
Constants as they appear in infinite volume. Therefore we can extract
physically relevant information even from the unphysical $\epsilon$-regime.

As a peculiarity of the $\epsilon$-regime, the topology plays an important
r\^{o}le. Observables tend to depend significantly on the topological sector,
and predictions exist for expectation values in specific sectors \cite{LeuSmi}.
%Therefore, numerical measurements should disentangle the topologies,
%: summing them up would be a drastic loss of information.
%in order to extract maximal information.
Therefore, for numerical measurements it can be advantageous, if not
necessary, to disentangle the topologies, in order to extract maximal
information.

In general, it is not obvious how to define topological sectors 
on the lattice, since all gauge configurations can be continuously 
deformed into one another. A neat definition exists, however, for
Ginsparg-Wilson fermions, since they have exact zero modes with definite
chiralities \cite{Has}. Once a GW operator is chosen, it has for each
gauge configuration a well-defined fermion index
\begin{equation}  \label{index}
\nu = n_{+} - n_{-} \ ,
\end{equation}
where $n_{\pm}$ is the number of zero modes with positive/negative
chirality. In the spirit of the Atiyah-Singer Theorem one then
uses this index as a definition of the topological charge.
For independent configurations the distribution of $\nu$ is Gaussian,
and its width determines the topological susceptibility.

Let us assume that we monitor the deformation of a gauge configuration 
to a different topological charge $\nu$ --- defined by the overlap 
operator given in eq.\ (\ref{overlap}). In the (real) spectrum of the
operator $Q$ such a change of $\nu$ means that some eigenvalue
crosses zero ({\em on} the topological boundary the overlap operator 
is not defined). 
For a transition the Monte Carlo history has to pass through a region
%in configuration space 
of rather low probability, and in simulations 
one assumes this to happen quickly at some instances in a long history. 
%The transitions occur in configurations of
%a rather low probability, and in Monte Carlo simulations one assumes it
%to happen quickly at some instances in a long history. 
For simulations in the $p$-regime one tries to sample
all topologies and frequent transitions are therefore desired.
However, in $\epsilon$-regime simulations one would often like to
collect statistics at one specific value of $| \nu |$ to measure
an expectation value in this sector. 
For the parameters that have been used in the $\epsilon$-regime simulations
\cite{sasa,spec,toposus,AA,zeromode,LMA,BS,Japeps,WenWit}, 
it would be of particular interest 
to collect large sets of configurations with $\vert \nu \vert =1$ and $2$.
The topologically neutral sector is problematic due to the frequent
appearance of very small Dirac eigenvalues, which leads to strong
spikes in the Monte Carlo histories of correlation functions \cite{AA}
(at $\vert \nu \vert > 0$ the non-zero eigenvalues are pushed to
higher energies \cite{DamNish,spec}). However, a procedure called 
Low Mode Averaging was designed to render also the neutral sector tractable 
\cite{LMA}, hence also a cumulation of configurations with $\nu =0$ may 
be of interest (see also Ref.\ \cite{Japeps}). 
Charged sectors are required, however, for the
method of extracting Low Energy Constants solely from the zero mode
correlation functions \cite{zeromode,BS}.

A box with $V \approx 10 ~{\rm fm}^{4}$ is suitable, but the width of 
the Gaussian charge distribution is then around 
$\langle \nu^{2} \rangle \approx 10$ \cite{toposus,BS}, 
so that most charges vary between about $-10$ and $10$. 
The index measurement by itself is computationally
expensive, hence identifying a set of, say, ${\cal O}(1000)$ configurations
in one sector is a tedious task --- if one uses the standard Wilson gauge 
action.

Therefore it is motivated to modify the lattice gauge action such
that topological transitions are suppressed.
\footnote{Also the use of multi-plaquette gauge actions, such as the
those suggested in Refs.\ \cite{PW,LW,Iwa,DBW2,FPA}, has some impact
on the frequency of topological transitions, see e.g.\ Refs.\ \cite{nonplaq}.
But such actions are less convenient for a systematic suppression of
small plaquette values, so we do not consider them here.} 
The vicinity of a transition
(under a continuous deformation) is characterised again by the occurrence
of small plaquette values, hence the technical aim is also in that
respect to avoid these configurations. This connection was
made rigorous first in Ref.\ \cite{HJL}.
If we consider the overlap operator (\ref{overlap}), 
the square root cannot vanish --- and therefore
topological transitions are excluded under continuous deformations ---
if all the plaquette variables $U_{P}$ in the configurations
involved obey the inequality (at $\mu =1$)
\begin{equation}  \label{epsineq}
S_{P}(U_{P}) % \equiv 1 - \frac{1}{3} {\rm Re~Tr}(U_{P}) 
< \varepsilon = \frac{2}{5d(d-1)} = \frac{1}{30} \ ,
\end{equation}
where $S_{P}$ represents the standard gauge action of one plaquette,
as specified in eq.\ (\ref{Wilact}).
Then the topological structure is continuum-like.

Later on H.\ Neuberger showed that this constraint can be relaxed
a little by increasing the threshold in inequality (\ref{epsineq}) to
\cite{Neu2}
\begin{equation}  \label{epsineqNeu}
\varepsilon = \frac{1}{(1 + 1/\sqrt{2})d(d-1)} \simeq \frac{1}{20.5} \ .
\end{equation}
This constraint ensures that the spectrum of $Q^{2}$ (the argument
of the square root in the overlap formula (\ref{overlap})) 
is strictly positive. 
\footnote{The corresponding admissibility condition has also been
studied on a non-commutative torus in Ref.\ \cite{Nagao}.}
%If this spectrum is above some constant larger 
%than zero, then the inverse square root can be expanded in a convergent
%power series, which also guarantees the locality of the overlap operator
In absence of zero modes this condition also guarantees the 
locality of the overlap operator (in the sense of an exponential 
decay) \cite{HJL}. 

%With respect to the topological
%charge one could object that a Monte Carlo history proceeds in discrete
%jumps, so even with this constraint the charge conservation is not absolutely 
%safe. However, this seems like a minor problem, since the charge would still be 
%conserved over very long periods in the history, and in simulations of
%the Hybrid Monte Carlo (HMC) type the few remaining changes could be 
%further suppressed by reducing the step size $d \tau $.

On the other hand, the impact of
imposing such a restrictive constraint strictly could be
a severe practical problem. The fluctuations of the gauge field 
would be limited so much that one could only obtain a tiny physical 
lattice spacing, and therefore a tiny physical volume. However, even
for simulations in the $\epsilon$-regime we have to require that 
the spatial box length $L$ exceeds some lower limit in the range 
of $L \gsim 1.1~{\rm fm} \dots 1.5~{\rm fm}$ (depending on the exact
criterion) \cite{spec,toposus,AA,zeromode,LMA,BS,Japeps}.

Here we present numerical experiments with gauge actions which do suppress
small plaquette values, but only to the extent that still allows for a
reasonable physical lattice spacing to be obtained. 
Then there is no rigorous guarantee for topology 
conservation in the Monte Carlo history.
\footnote{With respect to the topological
charge one could object that a Monte Carlo history proceeds in discrete
jumps, so even with this constraint the charge conservation is not absolutely 
safe. However, this seems like a minor problem, since the charge would still be 
conserved over very long periods in the history, and in simulations of
the HMC type the few remaining changes could be 
further suppressed by reducing the step size $d \tau $ 
(at higher cost, however).}
%--- not even under continuous 
%deformations, which corresponds to the limit $d \tau \to 0$. 
The hope is that
the transitions are still strongly suppressed, so that the history has
periods of constant charge, which are sufficiently long to allow us to
collect many configurations in a specific sector. Moreover, if we can be confident
that topological transitions rarely happen, most of the index 
computation can be omitted; one would just check after a number of 
configurations if the index has not changed.

Of course, these configurations 
should sample independently the observables to be measured in a fixed
topology. 
%This means that we aim at a {\em topological autocorrelation
%in the Monte Carlo history, which is much longer than the autocorrelation
%times with respect to other observables.} 
Since we aim at long sequences of fixed topological charge, this can 
also be interpreted as a long topological autocorrelation time. Of 
course, at the same time, we aim at a much shorter autocorrelation time 
for other observables.
We repeat that a long topological
autocorrelation is something one would not wish in general --- for instance
simulations in the $p$-regime --- but in the $\epsilon$-regime
this property turns into an advantage, if the observables of
interest are still weakly autocorrelated.

\section{The gauge actions}

We now describe a number of non-standard lattice gauge actions,
which suppress the undesired small plaquette values.
As long as the action for very smooth configurations ---
with $S_{P} \gsim 0$ for all plaquettes ---
is not altered, the naive continuum limit  coincides with
the one of the Wilson action (\ref{Wilact}), and with continuum
QCD. The suppression becomes strong when $S_{P}$ reaches the value of some 
parameter $\varepsilon$, which one would theoretically choose according 
to eq.\ (\ref{epsineqNeu}). For practical purposes
we will have to relax $\varepsilon$ to larger values.
A simple cutoff for $S_{P}$ at this value would be conceivable, but
such a discontinuity in the action (which would then suddenly jump
to infinity) does not appear promising. We can still impose a cutoff
but let the plaquette action diverge continuously as $S_{P}$ increases 
towards $\varepsilon$, if we modify $S_{P}$ of eq.\ (\ref{Wilact}) to
the hyperbolic form
\begin{equation}  \label{hypact}
S_{\varepsilon ,n}^{\rm hyp} (U_{P}) = 
\left\{ \begin{array}{cc}
\frac{S_{P}(U_{P})}{ [ 1 -  S_{P}(U_{P}) / \varepsilon ]^{n}} &
{\rm ~~for~~}  S_{P}(U_{P}) < \varepsilon \\
+ \infty & {\rm otherwise} \end{array} \right. 
\end{equation}
for $n > 0$. This formulation, with $n=1$, was introduced by M.\ L\"{u}scher
and used for conceptual studies of chiral gauge theories on the lattice
\cite{luschact}. In that case, $\varepsilon$ was of course set to a 
theoretically stringent value. 

In simulations, this action was first used in the Schwinger model by
H.\ Fukaya and T.\ Onogi \cite{FuOn}. They set $\varepsilon =1$,
i.e.\ far above the theoretical value of about $0.29$, but they still
observed topological stability over hundreds of trajectories.
\footnote{Note that the factor $1/3$ in the term for $S_{P}$ of 
eq.\ (\ref{Wilact}) is actually $1/N_{c}$ for general $SU(N_{c})$ 
or $U(N_{c})$ gauge groups. Moreover, the theoretical bound for 
$\varepsilon$ in $d=2$ is a factor $6$ larger than in $d=4$, as eqs.\
(\ref{epsineq}) and (\ref{epsineqNeu}) show
(this factor is due to a summation $\sum_{\mu > \nu}$ in the term
for $1/\varepsilon$).}

A theoretical objection against this lattice action was raised by
M.\ Creutz \cite{Creutz}. He observed that it does not provide
a positive definite transfer matrix. Of course, if we rely on the assumption 
that we are in the same universality class as the Wilson action
(as the naive continuum limit suggests) then one would not worry about that,
since the non-positivity would be a cut-off effect 
and we expect positivity to be restored in the continuum limit.
Still, a problem with this action cannot be strictly excluded.
Therefore we kept this point in mind in the numerical study and we will
comment on it in Subsection 4.3.

%For convenience one might consider using the hyperbolic part of
%the action (\ref{hypact}) for all plaquette values.
%A problem that would then plague practical applications is the scenario that
%in some HMC step the plaquette condition is ``jumped over'' for one or a few
%plaquettes. Then the potential for $S_{P}$ in the action (\ref{hypact})
%drives the following steps into a wrong direction,
%and the rest of the HMC history is completely distorted. Therefore this
%event must be excluded, i.e.\ we have to reject steps with such jumps
%and use action (\ref{hypact}) as it stands.

The infinite part in action (\ref{hypact}) means that certain steps
that the HMC algorithm suggests have to be rejected for sure. Therefore 
we also have to verify with special care that the acceptance rate is 
sufficiently high.

This motivated us to consider further variants of gauge actions,
which also suppress the probability of plaquette actions 
$S_{P} > \varepsilon$,
but which do not render a violation of this constraint completely
impossible. Examples for such actions are the ``power actions''and the
``exponential actions'',
\begin{eqnarray}  \label{powact}
S_{\varepsilon ,n}^{\rm pow} (U_{P}) & = & S_{P}(U_{P})
+ \frac{1}{\varepsilon} S_{P}(U_{P})^{n} \ , \\
S_{\varepsilon ,n}^{\rm exp} (U_{P}) & = & S_{P}(U_{P}) \cdot
\exp \{ S_{P}(U_{P})^{n} / \varepsilon \} \qquad ( n > 0 ) \ .
\label{expact}
\end{eqnarray}
%In these cases the exclusion of certain HMC steps is not needed.

In our numerical studies we included the actions
$S_{\varepsilon ,1}^{\rm hyp}$ and $S_{\varepsilon ,8}^{\rm exp}$.
Our preliminary results were reported in Refs.\ \cite{prelim}
and a comprehensive presentation will be given in the next Section.
Further results along these lines
for quenched QCD can be found in Refs.\ \cite{JapQCD,JapQCD2}.

\section{Numerical results}

\begin{figure}[htb]
\begin{center}
\includegraphics[angle=270,scale=0.48]{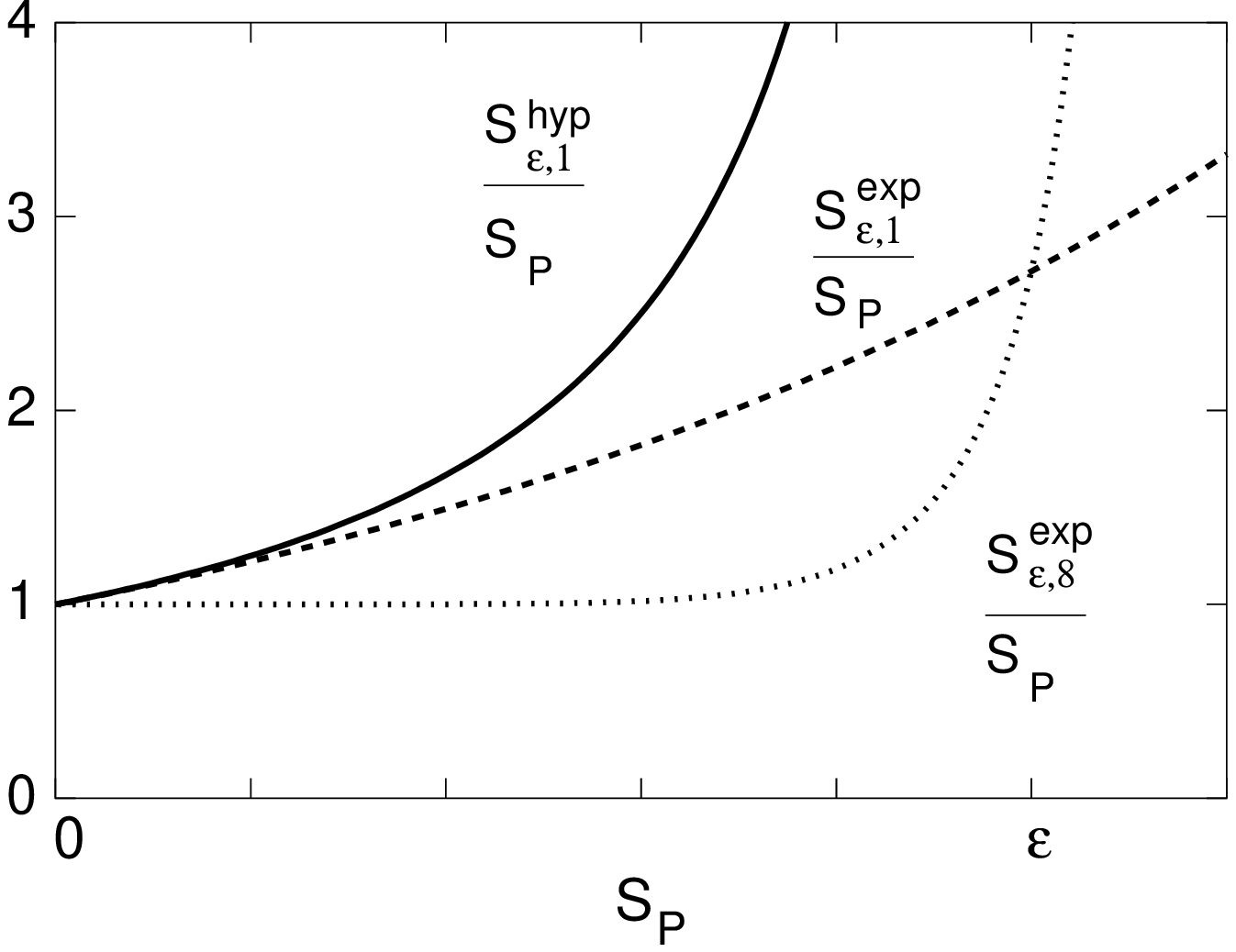}
\includegraphics[angle=270,scale=0.48]{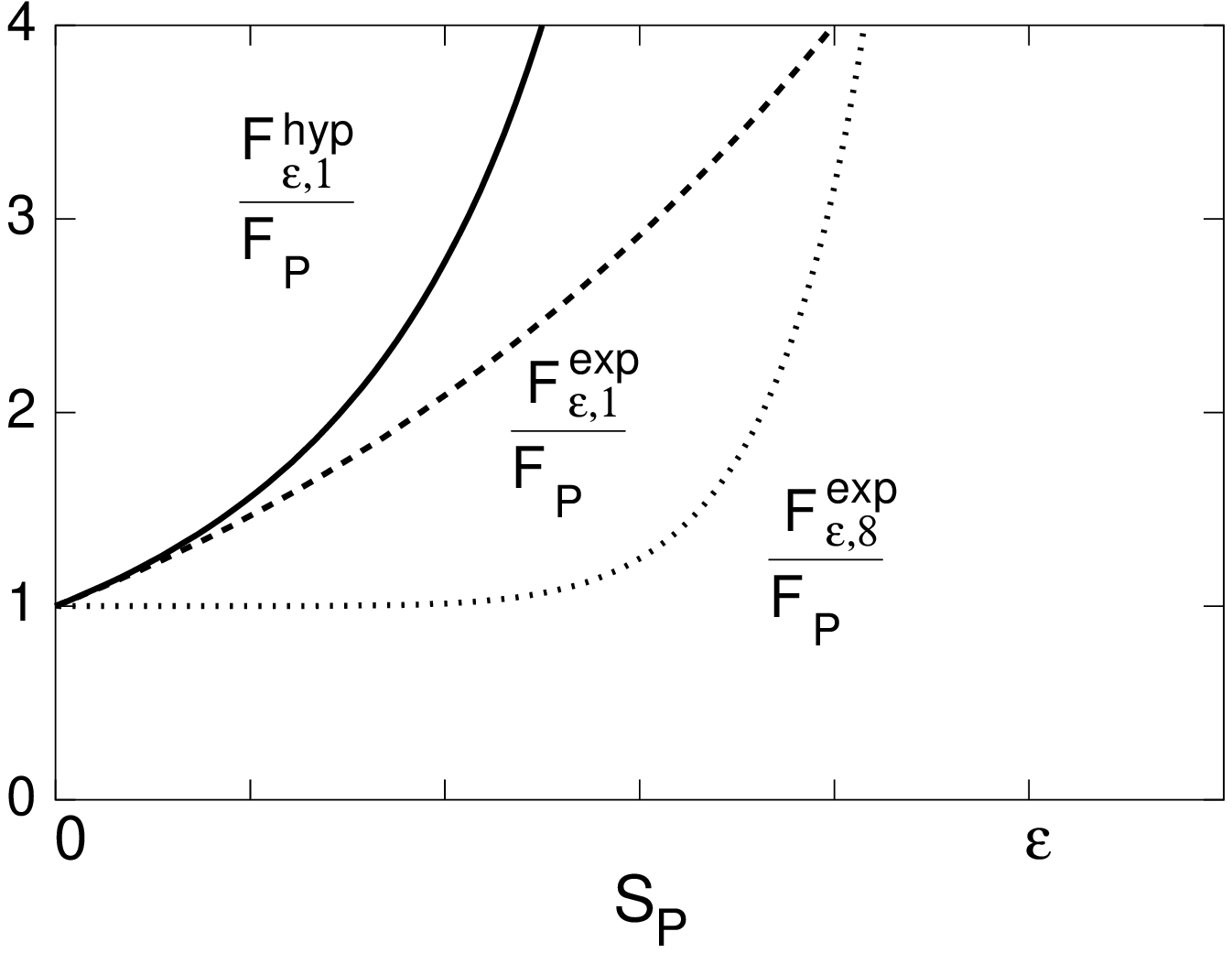}
\end{center}
\caption{\it On the left: 
The ratio between the hyperbolic plaquette action (for $n=1$)
and standard plaquette action, and the corresponding ratio for the exponential
actions with $n=1$ and $n=8$ (the latter is that case we studied). 
On the right: the same ratios for the HMC forces, again as a function
of $S_{P}$.}
\label{actfor}
\end{figure}

Actions of the types (\ref{hypact}), (\ref{powact}) and
(\ref{expact}) depend non-linearly on the link variables $U_{x,\mu}$.
As a consequence, the heat-bath algorithm and over-relaxation
cannot be applied straightforwardly.
Therefore we use the {\em local HMC algorithm}, which was introduced 
in Ref.\ \cite{lHMC}. Since these actions are still composed of separate
contributions by the single plaquettes, the force in the local HMC algorithm
is a simple modification of the corresponding force for the Wilson
action,
\begin{eqnarray}
\hspace*{-16mm} && 
F^{\rm hyp}_{\varepsilon , n} = 
\frac{\delta S^{\rm hyp}_{\varepsilon , n} (U_{P})}{\delta U_{x,\mu}} 
= F^{\rm W} (U_{P}) \cdot
\frac{1 + \frac{n-1}{\varepsilon} S_{P}(U_{P})}
{[1 - \frac{1}{\varepsilon} S_{P}(U_{P})]^{n+1}} \\
\hspace*{-16mm} &&
F^{\rm exp}_{\varepsilon ,n} = 
\frac{\delta S^{\rm exp}_{\varepsilon ,n} (U_{P})}{\delta U_{x,\mu}} 
=   F^{\rm W} (U_{P}) \cdot
\Big( 1 + \frac{n}{\varepsilon} S_{P}(U_{P})^{n-1} \Big) \,
\exp \Big\{ \frac{1}{\varepsilon} S_{P}(U_{P})^{n} \Big\} \\
\hspace*{-16mm} && {\rm where} \qquad
 F^{\rm W} (U_{P}) = \frac{\delta S_{P} (U_{P})}{\delta U_{x,\mu}}
\nonumber
\end{eqnarray}
is the force of the Wilson action, and its modification by the
second factor is of order ${\cal O}(S_{P}/\varepsilon )$ in both cases.  
This is illustrated in Figure \ref{actfor}. \\

In such simulations, it is also of special importance to check that the
results do not depend on the starting configuration (once we start in the
desired topological sector). It would be conceivable that the constraint
on the plaquette values causes also unwanted obstructions. For all
the quantities to be considered below, it turned out that this was
not the case; an example is discussed in Subsection 4.6.

\subsection{Plaquette values}

\begin{figure} 
\vspace*{-1cm} 
\begin{center}
  \includegraphics[angle=0,width=.75\linewidth]{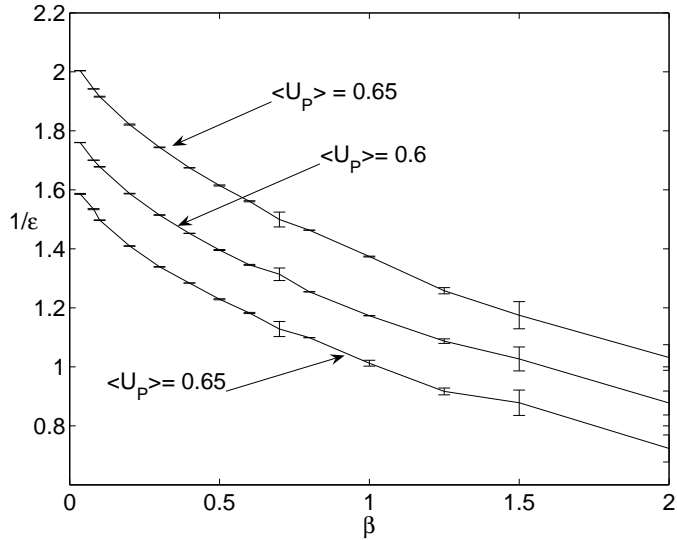}
\end{center}
  \caption{{\it The lines of constant plaquette values in the
plane spanned by $1/ \varepsilon$ and $\beta$, % evaluated 
on a $4^{4}$ lattice.}}
\label{conspla}
\end{figure}

As a first experiment we considered action (\ref{hypact}) and
searched for the lines of a constant mean plaquette action $\langle S_P 
\rangle$ on a $4^{4}$
lattice, as $\beta $ and the action parameter $\varepsilon$ are varied.
The result is shown in Figure \ref{conspla}. As we decrease $\varepsilon$,
very small values of $\beta$, i.e.\ strong bare gauge couplings 
are needed to keep $\langle S_P \rangle$ constant.

%\vspace*{-3cm}

\begin{figure} 
\vspace*{-1cm} 
\begin{center}
  \includegraphics[angle=0,width=.58\linewidth]{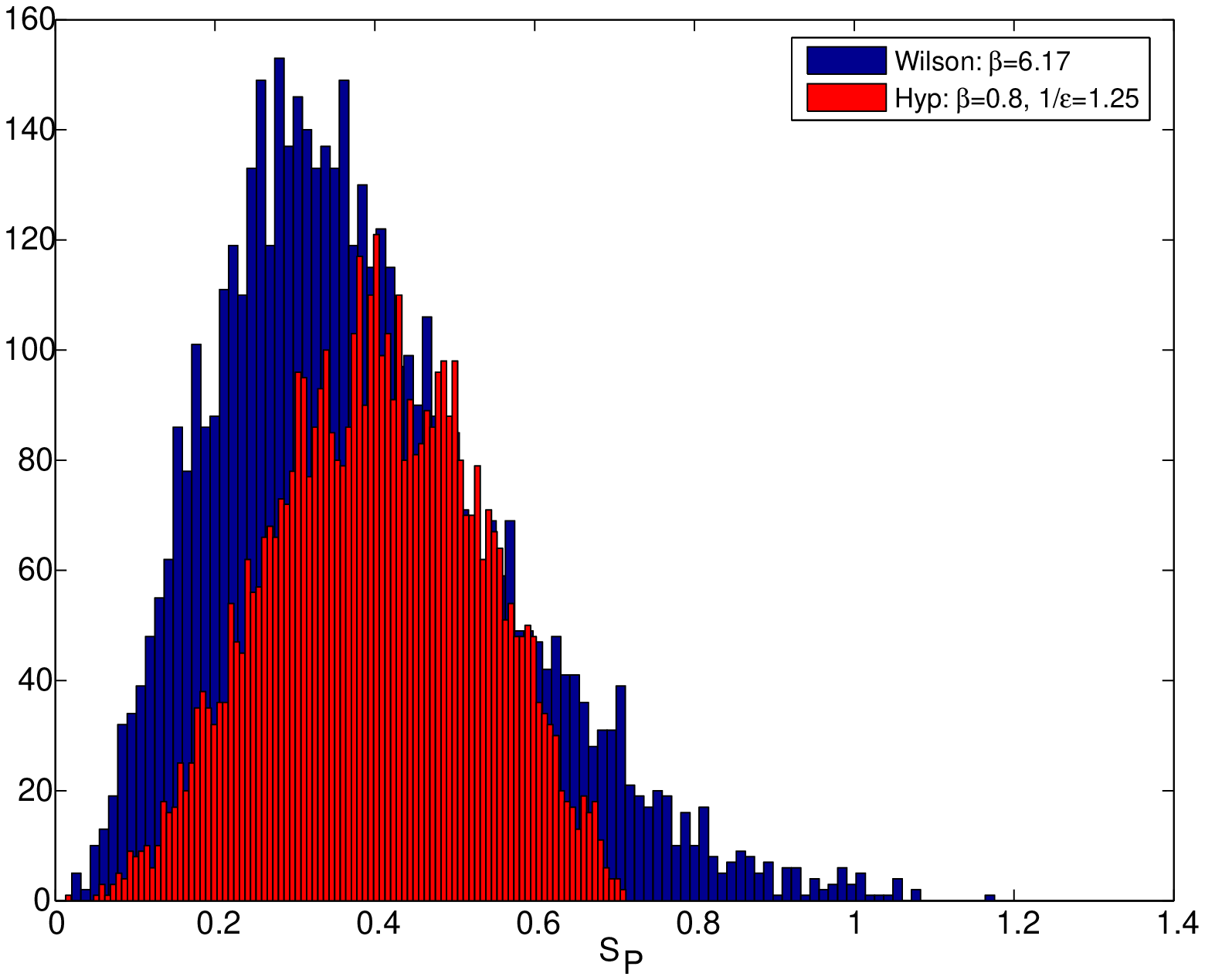}
  \includegraphics[angle=0,width=.58\linewidth]{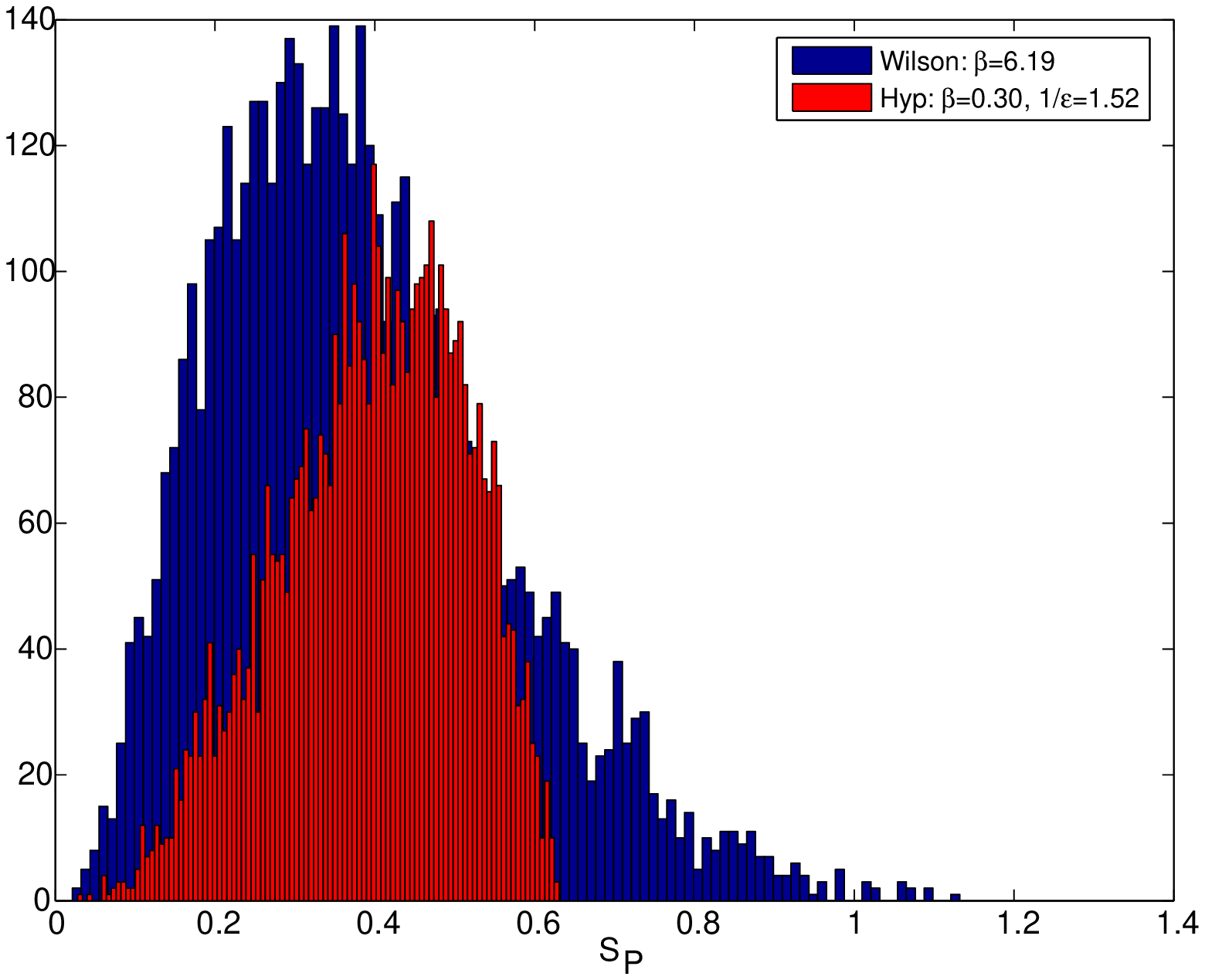}
  \includegraphics[angle=0,width=.58\linewidth]{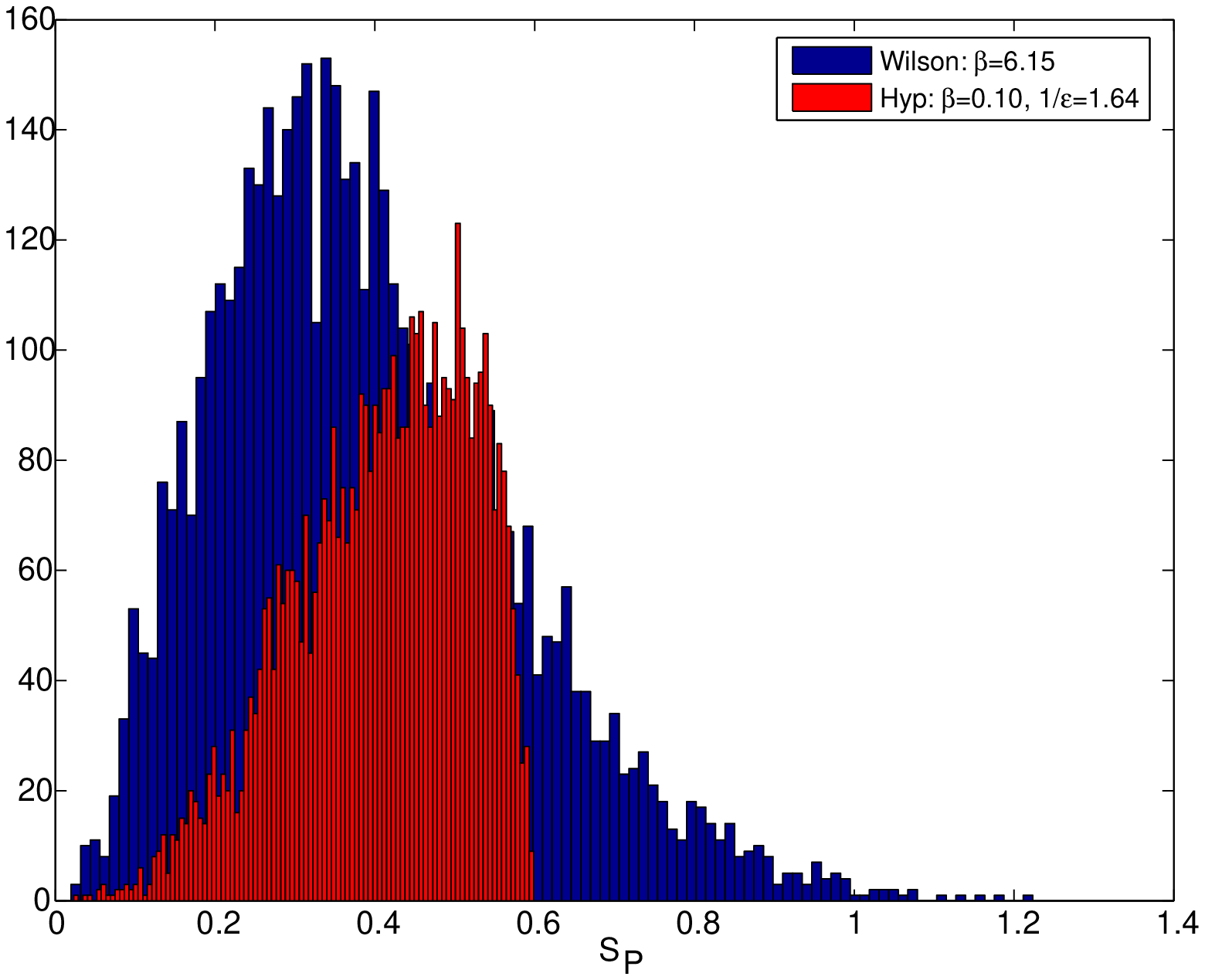}
\end{center}
\caption{{\it Examples for the histograms of plaquette actions 
$S_{P}(U_{P})$ with hyperbolic topology conserving actions
$S_{\varepsilon ,1}^{\rm hyp}$ at various parameters,
compared to the Wilson action, on a $4^{4}$ lattice.}}
\label{plaqhisto}
\end{figure}

Next we took a look at the statistical distribution of
the plaquette actions,
and we show corresponding histograms in Figure \ref{plaqhisto}.
By decreasing the values of $\varepsilon$ and $\beta$
we can in fact keep $\langle S_{P} \rangle$ approximately constant, while
drastically suppressing the occurrence of very small plaquette actions.

However, for a systematic approach to identify a line of a constant physical 
scale, we have to proceed to larger lattices and we determine the physical
scale based on the static potential, since
the plaquette value cannot be used for this purpose.
This issue will be discussed in the next Subsection.

\subsection{The static potential and the physical scale}

A very well established method for setting a scale in pure gauge theory
is based on the measurement of the static potential at intermediate
distances.
This potential, and the resulting force, are extracted from the Wilson
loop correlations at sufficiently large time separations, such that 
excited states can be neglected. 

To suppress the effects of short-ranged fluctuations, we applied APE smearing
\cite{APE} on the spatial links before evaluating the Wilson loops. 
Then we evaluated the static potential
\begin{equation} \label{potential}
a V(r) = \ln \left( \frac{W(r,t-a)}{W(r,t)} \right) \ ,
\end{equation}
where $W(r,t)$ is a rectangular Wilson loop with extension $r$ and $t$
in a spatial and in the temporal direction 
(and $V(r)$ is extracted from an asymptotic window in $t$).
Here we applied the 
variational procedure explained in Refs.\ \cite{GSW,NeSo}.
The force $F(r)$ is then obtained from the discrete gradient of $V(r)$.
The aim is to fix the quantity $r_{0}=0.5 ~{\rm fm}$ \cite{RSo} 
by tuning the dimensionless term
\begin{equation}
r^{2} F(r) \vert_{r= r(c)} = c \ , \quad r_{0} = r (1.65) \ .
\end{equation}
To this end, we applied a local interpolation for the force based 
on the ansatz
\begin{equation} \label{interpol}
F(r) = f_{1} + f_{2} r^{2} \ .
\end{equation}
The results for the action $S^{\rm hyp}_{\varepsilon,1}$ at different 
values of $\varepsilon$ and $\beta$ are shown in Table 1. %\ref{tab_r0_1}.
From the scale $r_0/a$ we can identify an ``equivalent'' $\beta$-value
for the Wilson gauge action, which we denote by
$\beta_{\rm W}$. It corresponds to the same lattice spacing for the
plaquette action (\ref{Wilact}), where we refer to the 
parametrisation formula in Ref.\ \cite{GSW}.

\begin{table}
%\begin{center}
\hspace*{-1cm}
\begin{tabular}{|c|c|c|c|c|c|c|c|c|}
\hline
$1 / \varepsilon $  & $\beta$ & $r_0/a$ & $\beta_W$ & $d \tau$ 
& $\tau^{\rm plaq}$ & $f_{\rm top}$ & $\tau^{\rm plaq} \cdot f_{\rm top}$ 
& acc.\ rate \\
\hline
\hline
0. & 6.18 & 7.14(3) & 6.18 & 0.1    & 7(1)   & 2.2(13) e-2 & $\approx$ 1.5 e-1 & 
$>$ 99 \% \\
\hline
1. & 1.5 & 6.6(2) & 6.13(2) & 0.1   & 2.2(1) & 3.0(23) e-3 & $\approx $ 6.6 e-3
& $>$ 99 \% \\
1. & 1.5 & 6.6(2) & 6.13(2) & 0.05  & 2.0(1) & 2.9(11) e-3 & $\approx $ 5.8 e-3
& $>$ 99 \% \\
1. & 1.5 & 6.6(2) & 6.13(2) & 0.01  & 2.2(1) & 3.5(8)  e-3 & $\approx $ 7.7 e-3
& $>$ 99 \% \\
1. & 1.5 & 6.6(2) & 6.13(2) & 0.005 & 2.3(2) & 2.8(15) e-3 & $\approx $ 6.4 e-3
& $>$ 99 \% \\
\hline
1.18 & 1. & 7.2(2) & 6.18(2) & 0.1  & 1.2(1) & 2.0(12) e-3 & $\approx $ 2.4 e-3
& $>$ 99 \% \\
1.18 & 1. & 7.2(2) & 6.18(2) & 0.02/0.01 & 1.3(1) & 1.6(7) e-3 
& $\approx $ 2.1 e-3 & $>$ 99 \% \\
\hline
1.25 & 0.8 & 7.0(1) & 6.17(1) & 0.1 & 1.1(1) & 2.3(13) e-3 & $\approx $ 2.5 e-3
& $>$ 99 \% \\
\hline
1.52 & 0.3 & 7.3(4) & 6.19(4) & 0.1 & 0.8(1) & 9.0(28) e-4 & $\approx $ 7.2 e-4
& $\approx$ 95 \% \\
\hline
1.64 & 0.1 & 6.8(3) & 6.15(3) & 0.1 & 1.0(1) & 1.3(7) e-3 & $\approx $ 1.3 e-3
& $\approx$ 65 \% \\
\hline
1.64 & 0.1 &        &         & 0.05 & 0.7(1) & 2.3(13) e-3 & $\approx $ 1.6 e-3
& $\approx$ 78 \% \\
\hline
1.64 & 0.1 &        &        & 0.025 & 0.6(1) & 3.5(20) e-3 & $\approx $ 2.1 e-3
& $\approx$ 93 \% \\
\hline
1.64 & 0.1 &        &        & 0.001 & 0.5(1) & 3.7(23) e-3 & $\approx $ 1.9 e-3
& $\approx$ 99 \% \\
\hline
\end{tabular}
\vspace*{3mm}
%\end{center}
\caption{\it Results for the hyperbolic actions $S^{\rm hyp}_{\varepsilon ,1}$,
defined in eq.\ (\ref{hypact}), for different values of $\varepsilon$ and 
$\beta$, on a $16^{4}$ lattice. 
We first show the ratio $r_0/a$, which fixes the physical scale.
For comparison we also display the $\beta$-values
$\beta_{\rm W}$, which leads to the same physical scale for the Wilson 
action (\ref{Wilact}) \cite{GSW}. 
The trajectories were all of length 1 and 
divided into HMC steps of length $d \tau$. For the plaquette values this leads
to a mean autocorrelation time  $\tau^{\rm plaq}$, which we show as an
example for the autocorrelation of a non-topological quantity.
The topological stability, on the other hand, is measured by the 
frequency of topological transitions. More precisely, $f_{\rm top}$ is 
the  number of topological jumps (determined from cooling),
normalised by the number of trajectories.
Its product with $\tau^{\rm plaq}$ characterises the dominance of the
topological autocorrelation.
Finally we give the acceptance rate of the local
HMC algorithm. For each set of parameters in this Table we collected
at least 200 thermalised configurations spaced by 50 trajectories each.
A detailed discussion is given in the following Subsections.}
\label{tab_hyp}
\end{table}

\begin{table}
\hspace*{-1cm}
%\begin{center}
\begin{tabular}{|c|c|c|c|c|c|c|c|c|}
\hline
$1 / \varepsilon$ & $\beta$ & $r_0/a$ & $\beta_W$ & $d \tau$ 
& $\tau^{\rm plaq}$ & $f_{\rm top}$ & $\tau^{\rm plaq} \cdot f_{\rm top}$ &
acc.\ rate \\
\hline
\hline
%1 & 5.6  & 0.1       &        &         & 0.1   & 1.3(1)   & 4.1e-3 & \\
%\hline
%2 &  9   & 0.1       &        &         & 0.1   & 0.7(1)   & 5.8e-3 & \\
%\hline
%3 & 15   & 0.1       &        &         & 0.1   & 0.7(1)   & 4.6e-3 & \\
%\hline
%4 & 30   & 0.1       &        &         & 0.1   & 0.7(1)   & 9e-4   & \\
%\hline
%6 & 90   & 0.1       &        &         & 0.1   & 0.7(1)   & 8e-4   & \\
%\hline
 500 & 0.044 & $>8.5$ & $>6.30$ & 0.015 & 0.6(1)  & 2.5(36) e-4 & 
$\approx$ 1.5 e-4 & $\approx$ 99\%\\
\hline
 600 & 0.0134 & 8.0(2) & 6.26(1) & 0.015 & 0.6(1) & 2.5(23) e-4 & 
$\approx$ 1.5 e-4 & $\approx$ 99\%\\
\hline
 1000 & 0.004 & $>9$ & $> 6.34$ & 0.03 & 0.6(1)   & 0(0) & & $ \approx$ 25\%\\
\hline
 1000 & 0.00113 & 7.9(1) & 6.25(1) & 0.015 & 0.6(1) & 1.7(23) e-4 &
$\approx$ 1.0 e-4 & $\approx$ 99\%\\
\hline
\end{tabular}
%\end{center}
\vspace*{3mm}
\caption{\it Results for the exponential actions $S^{\rm exp}_{\varepsilon,8}$,
defined in eq.\ (\ref{expact}), for different values of $\varepsilon$
and $\beta$, on a $16^{4}$ lattice. 
As in Table 1 we first show the ratio $r_0/a$ and the Wilson $\beta$-value
$\beta_{\rm W}$, which leads to the same physical scale for the Wilson 
action (\ref{Wilact})
(here the finite size effects in the evaluation of $r_0/a$ may be sizable).
For different HMC steps $d \tau$ we then give 
the mean plaquette autocorrelation time  $\tau^{\rm plaq}$,
the frequency of topological transitions, $f_{\rm top}$, its product
with $\tau^{\rm plaq}$ and
the acceptance rate. The number of measurements for $r_{0}/a$ was at least
200 in each case. Further comments are added in Subsections 4.4 and 4.6.}
\label{tab_exp}
\end{table}

Let us comment now on the possible errors in this evaluation:

\begin{itemize}

\item The systematic {\em error on the interpolation} can be estimated 
by considering a third point and observe the relative deviation between 
the two interpolations; in our case this error turned out to be negligible 
compared to the statistical uncertainties.

\item Previous computations \cite{GSW} revealed 
that for a box size $L\gtrsim 3.3\;r_0$ the {\em finite size effects} 
for the computation of $r_0$ can be safely neglected.
The same work observed that for the Wilson action at $\beta=5.95$
and $L\simeq 2.4 \, r_0$ the finite size artifacts of the force 
amount to $\approx \;3\%$.
In our study we deal with $L=16 \simeq (2.2 \dots 2.4)\;r_0$.
Hence we assume finite size effects for $r_0$ to
be of this order as well.

\item The errors quoted on $r_0/a$ in Table \ref{tab_hyp} and \ref{tab_exp}
are purely {\em statistical} (they were computed by the jackknife method).
The same errors are shown in Figures 4 and 5, which we comment on below.
% should be \ref{scalefig1} and \ref{scalefig2}, 
% but for some reason it doesn't work
The extent of these errors is acceptable in this
context, since a precise determination of $r_0$ 
was not the purpose of this work, so we did not aim at high statistics.

\item A way to check the {\em lattice artifacts} 
is to compare the short distance force at finite lattice spacing with the one
extrapolated to the continuum limit in Ref.\ \cite{NeSo}.
In particular we measured the ratio
\begin{equation}  \label{ratforce}
\Delta(r/r_0)=\frac{r^2 F(r/r_0)-r^2 F(r/r_0)|_c}{r^2 F(r/r_0)|_c} \ ,
\end{equation}
where $r^2 F(r/r_{0})|_c$ denotes the continuum limit.
The results for the action $S^{\rm hyp}_{\varepsilon,1}$ are shown in
Figure 4. % \ref{scalefig1} . 
At short distances, lattice artifacts are below $15\%$ for 
all the different values of $1/\varepsilon$ and $\beta$ that we included. 
Moreover, one observes that discretisation errors grow substantially for
increasing values of $1/\varepsilon$, as expected. This indicates
that choosing even larger values of $1/\varepsilon$ --- while keeping the 
physical lattice spacing fixed --- could introduce sizeable cutoff effects.   

\begin{figure}
\vspace*{-1cm}
\begin{center}
\includegraphics[width=10cm,angle=-90]{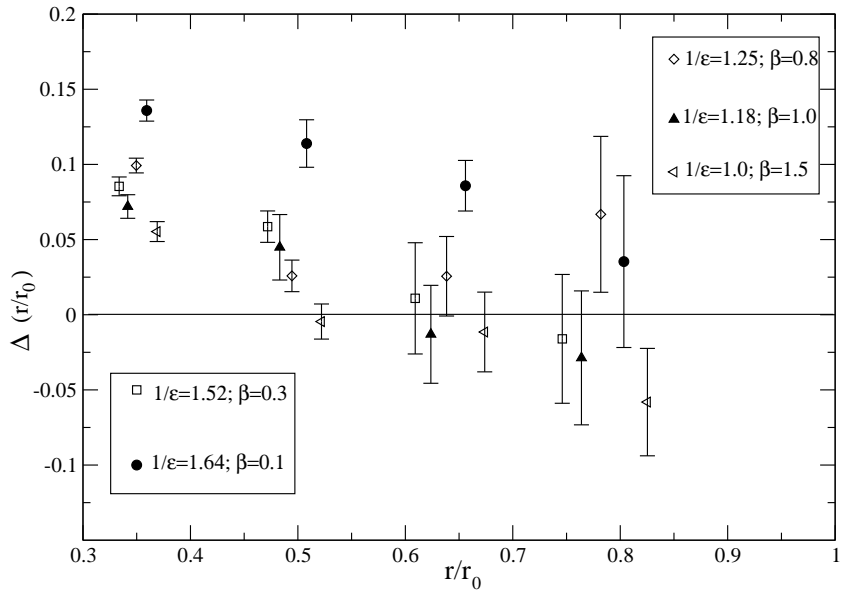}
\vspace{-0.7cm}
\caption{\it Lattice artifacts for the action $S^{\rm hyp}_{\varepsilon,1}$. 
The plot shows the relative deviation of the force $r^2F(r)$ from the 
continuum results, given in eq.\ (\ref{ratforce}).}
\end{center}
\vspace*{-7mm}
\label{scalefig1}
\end{figure}

\item For the Wilson action it turned out that the lattice artifacts can
be reduced considerably by using a so-called {\em ``tree
level improved''} definition of the force.
For this purpose, one defines an improved distance $r_I$ such that the force
does not contain lattice artifacts at tree level,
\begin{equation}  \label{impforce}
F(r_I)=\frac{4}{3}\frac{g_0^2}{4\pi r_I^2}+\mathcal{O}(g_0^4) \ .
\end{equation}
If we adapt this method, the procedure described before leads to
the results shown in Figure 5. %\ref{scalefig2}.
For the largest values of $1/\varepsilon$ one observes some reduction of the
lattice artifacts, whereas they seem to increase for the smallest $1/\varepsilon$.
This behaviour is not totally surprising, since it has been observed in other 
cases that this improvement is not always guaranteed \cite{SNe}.\\ 
We also observe that the use of tree-level improved observables does not 
change the results for $r_0/a$ itself (within the statistical errors). 

\item For a comparison of the scaling quality one may, for instance, refer to
the Iwasaki action \cite{Iwa} (at $r_0/a\simeq 6.0$) and the
DBW2 action \cite{DBW2} (at $r_0/a \simeq 5.5$) 
at a distance $r/r_0 \approx 0.3$: 
in these cases the lattice artifacts were found to be
of order $\sim 10\%$ \cite{SNe}. Those discretisation
errors are therefore comparable to our results for the actions
$S^{\rm hyp}_{\varepsilon,1}$.

\end{itemize}

\begin{figure}
\vspace*{-1cm}
\begin{center}
\includegraphics[width=10cm,angle=-90]{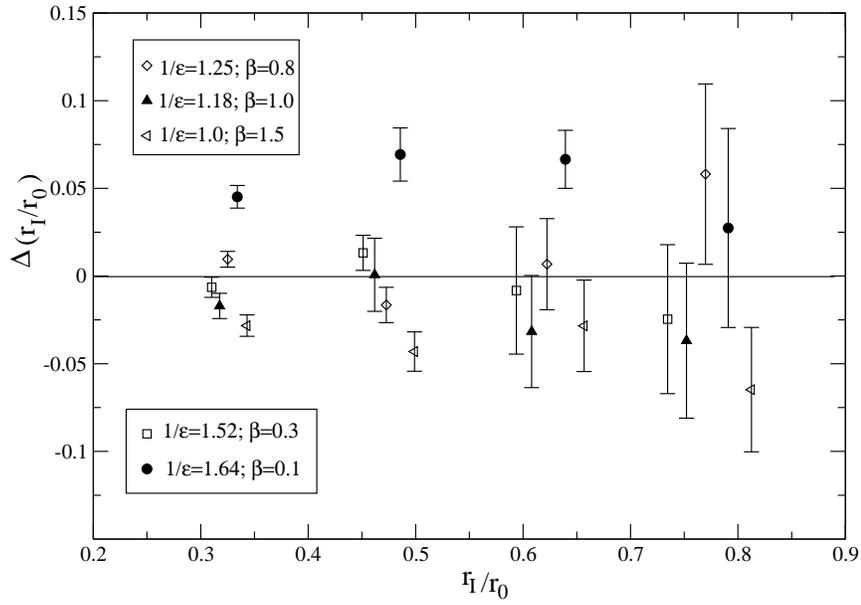}
\vspace{-0.7cm}
\caption{\it Lattice artifacts for the action $S^{\rm hyp}_{\varepsilon,1}$ 
with a tree level improved definition of the force (\ref{impforce}). The plot
shows the relative deviation of the force $r^2F(r)$ from the continuum 
results.}
%The symbols have the same meaning as in Figure 4.}
\end{center}
\label{scalefig2}
\end{figure}

We repeat that our statistics is modest, and our intention in this
analysis was only to check whether the errors and lattice artifacts 
are reasonably under control. This can be confirmed from the results 
shown Figures 4 and 5, and the accuracy is fully sufficient for our
purposes.

\subsection{The issue of reflection positivity}

As we anticipated in Section 3, Ref.\ \cite{Creutz} pointed out that
actions of the type (\ref{hypact}) do not provide a positive transfer 
matrix, as it was observed \cite{NoManton}
a long time ago for the so-called Manton action \cite{Manton}.
The positivity of the transfer matrix is assured if positivity holds 
for both, ``site reflection'' and ``link reflection'' (e.g.\ reflections 
on the hyperplanes $x_{4} = 0$ and $x_{4} = 1/2$).
%The positivity of the transfer matrix is assured if positivity holds
%for ``link reflections'', i.e.\ reflections on a hyperplane at a 
%half-integer time coordinate.

However, this still leaves the possibility of obtaining a positive
{\em squared} transfer matrix. In fact, this is the case if the action
obeys at least {\em one} of the two reflection positivities \cite{MoMu}. 
%In this case, one refers to a reflection at a Euclidean time slice, say
%$t =0$. 
We have checked for the actions involved in our study, given in eqs.\
(\ref{hypact}) and (\ref{expact}), that site reflection positivity
does in fact hold.

From the practical point of view, the lack of positivity can be
reflected in an irregular behaviour of the effective potential $V(r,t)$ 
at short time sepa\-ration, as observed in Ref.\ \cite{RSo} for actions 
involving rectangular loops (like those suggested in Refs.\
\cite{LW,Iwa,DBW2}). This is a lattice artifact which does, in principle,
not constitute a problem, as long as one keeps far enough from the cutoff.
\footnote{Problems can arise in the application of the variational
method, where one has to choose a small reference time.} In the case of
the actions considered in this work and with our statistical precision, 
such an irregular behaviour could not been observed in the extraction 
of the static potential. 

%Moreover, if an action is plagued by a conceptual problem with respect
%to the continuum limit, then one often observes an irregular behaviour
%in the short-distance force. Indeed, this was observed exactly for
%certain lattice gauge actions which lack positivity \cite{SNe}.
%However, in the case of the actions considered here, we see a consistently
%regular behaviour of the force at short distances, similar to other
%established gauge actions \cite{Wilson,LW,Iwa,DBW2}, and fully consistent 
%with the correct continuum limit.

\subsection{Acceptance rate}

Also the problem related to the acceptance rate has been mentioned in
Section 3. This point motivated us to consider also a set of
exponential actions of the type (\ref{expact}), in addition to
the hyperbolic actions, and the corresponding results are given
in Table \ref{tab_exp}. In both cases, the acceptance rate is
very high for most of the actions we studied. It drops, however,
if one pushes for very low values of $\varepsilon$ (along with
an extremely small $\beta$). As the last lines in Table \ref{tab_hyp}
show, the acceptance rate can actually be driven up again even at
$1 / \varepsilon = 1.64$ by using very short HMC steps. However, 
this cannot be considered a solution, because it increases the costs
(especially in the dynamical case), and also the frequency of topological 
transitions rises again (c.f.\ Subsection 4.6).

Therefore, this property sets another 
limit on the suppression of the small plaquette values, in addition
to the scaling of the static potential at short distances.

\subsection{The condition number for the overlap operator}

This Subsection discusses the condition number of the operator $Q^{2}$,
which is crucial for the computational effort required to deal with
the overlap operator in eq.\ (\ref{overlap}). Table \ref{cond_tab}
collects our results for the action $S^{\rm hyp}_{1,\varepsilon}$
compared to $S_{\rm W}$, at the values of $\varepsilon$ and $\beta$
corresponding to approximately constant physics according to Table 
\ref{tab_hyp}. Similar results were presented in Refs.\ \cite{JapQCD2},
which also include first tests with dynamical overlap fermions.

\begin{figure} 
%\vspace*{-1cm} 
\begin{center}
  \includegraphics[angle=270,width=.7\linewidth]{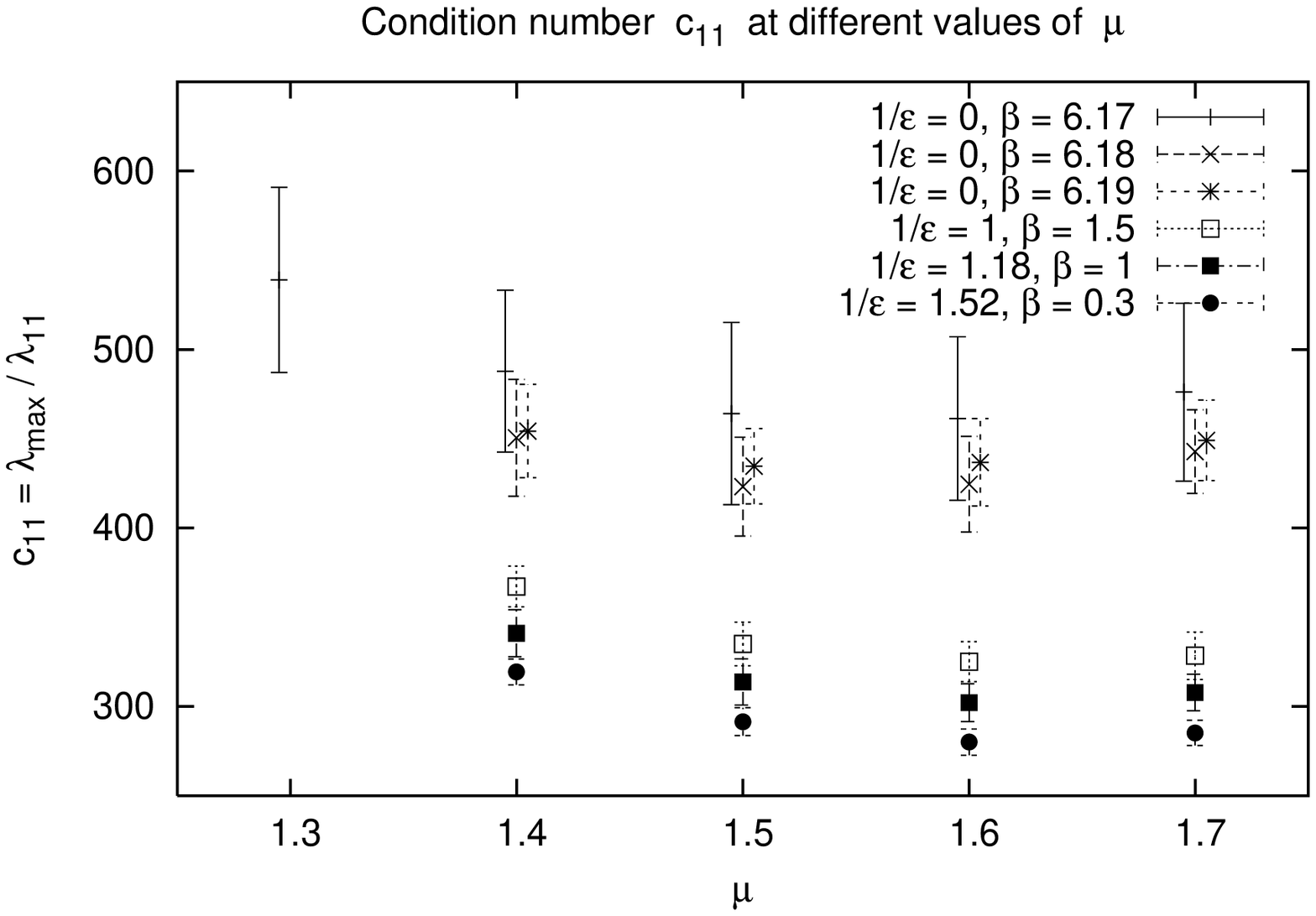}
  \includegraphics[angle=270,width=.7\linewidth]{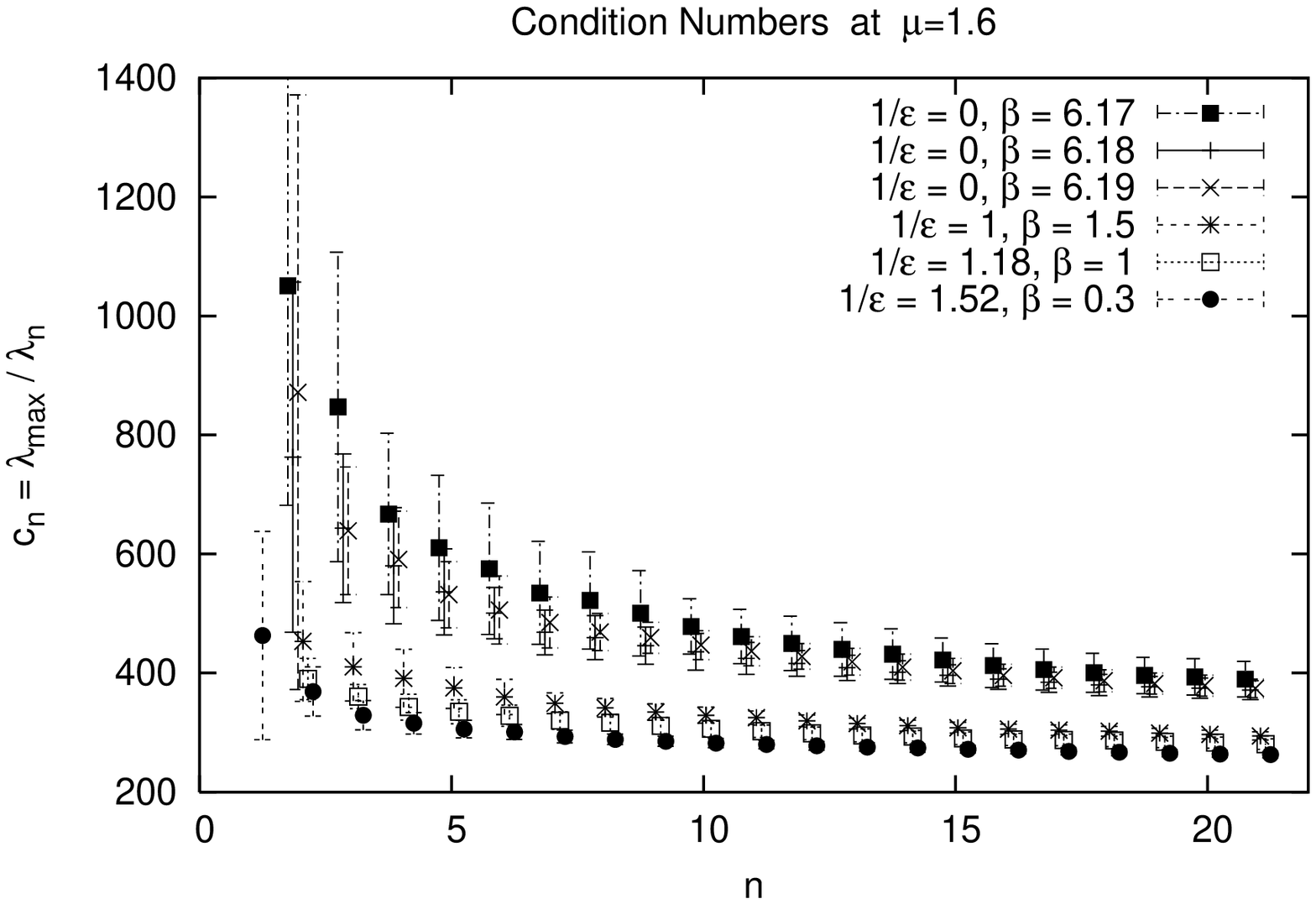}
\end{center}
\caption{{\it Comparison of the condition number $c_{11}$ 
for various values of $\mu$ (on top),
and $c_{n}$ for running $n$ at $\mu =1.6$ (below),
for different types of gauge actions.
(The parameters $\mu$ and $c_{n}$ are defined in eqs.\
(\ref{overlap}) and (\ref{cneq}).)}}
\label{condfig}
\end{figure}

In our study we varied the parameter $\mu$ by units
of $0.1$ and found the optimal condition numbers for all gauge actions
involved at $\mu =1.6$. For the case of 10 eigenmodes of $Q^{2}$
projected out, this property can be seen in the upper plot of Figure
 \ref{condfig}.
Hence we compare the condition numbers at $\mu =1.6$
for different gauge actions in Table \ref{cond_tab} and in
the lower plot of Figure \ref{condfig}. These results are
based on 30 configurations in each case.

\begin{table}
\begin{center}
\begin{tabular}{|c|c|c|c|c|c|}
\hline
$\beta$ & $1/ \varepsilon$  & $c_{2}$ & $c_{6}$ & $c_{11}$ & $c_{21}$ \\
\hline
\hline
6.17 &  0   & 1051(369) &  575(110) &  461(46)  &  390(30) \\
\hline
6.18 &  0   &  723(294) &  501(43)  &  424(27)  &  371(16) \\
\hline
6.19 &  0   &  872(499) &  506(57)  &  437(25)  &  374(16) \\    
\hline
1.5  &  1   &  453(101) &  360(30)  &  325(11)  &  294(7)  \\
\hline
 1   & 1.18 &  390(34)  &   328(18) &  302(10)  &  281(9)  \\
\hline
0.8  & 1.25 &   439(89) &   341(23) &  311(15)  &  285(9)  \\
\hline
0.3  & 1.52 &   369(41) &   301(13) &  280(7)   &  263(5)  \\
\hline
0.1  & 1.64 &   433(82) &   342(18) &  315(14)  &  293(8)  \\
\hline
\end{tabular}
\end{center}
\caption{\it Condition numbers $c_{n}$ of the operator $Q^{2}$
in the square root of the Neuberger overlap operator
at $\mu = 1.6$, after projecting out the leading $n-1$ modes of 
$Q^{2}$. In this comparison we always considered configurations
generated by the local HMC algorithm with $d \tau = 0.1$.}
\label{cond_tab}
\end{table}

We show the condition numbers
\begin{equation} \label{cneq}
c_{n} := \lambda_{\rm max} / \lambda_{n} \ , \quad 
(\lambda_{\rm max}, \ \lambda_{n} \ : \ {\rm largest~resp.~n^{th}
~eigenvalue~of~}Q^{2})
\end{equation}
which are relevant after projecting out the leading
$n-1$ modes of $Q^{2}$. 
\footnote{We have checked that the polynomial degree for a fixed
precision of the overlap operator is $\propto \sqrt{c_n}$
(to a high precision). At the side-line, we also observed that there
does not seem to be to any dependence of the condition numbers $c_{n}$ on the 
topological sector.}
We see that the $c_{n}$ are indeed
lowered as $1/ \varepsilon$ increases, which reduces the effort
for overlap fermion simulations. 
\footnote{Alternatively, lower condition numbers $c_{n}$ can also
be achieved by inserting an improved kernel $Q$ into the overlap 
formula, see for instance Refs.\ \cite{ovimp,BS}.} 
If $n$ is around 20, then ---
for the smooth configurations that we considered here ---
the gain compared to $S_{\rm W}$ is only moderate. 
However, for the hyperbolic actions
$S^{\rm hyp}_{1,\varepsilon}$ the number of these modes can be reduced
drastically without much loss in the condition number of the remaining 
operator. This is in contrast to $S_{\rm W}$, and it matters
in applications, since the special treatment of each of these 
projected modes also takes computation time
(although this is typically a minor part of the total computational
effort).
%Therefore the strongly
%reduced number of modes to be treated separately is the main gain
%factor that we observe here for the topology conserving gauge actions.

\subsection{Topological charge stability}

For a quick analysis, we used the cooling method \cite{cool}
to estimate the topological charge. The resulting topological
stability over the trajectories is included in Tables \ref{tab_hyp}
and \ref{tab_exp}. In independent tests we evaluated for a subset of the
configurations the overlap indices setting $\mu = 1.3$, $1.4$, $1.5$, $1.6$ 
and $1.7$. Since we are dealing with smooth 
configurations, it does not come as a surprise that we found an excellent 
agreement of more than 98 \% for all these definitions of
the topological charge, i.e.\ the charges obtained with
cooling and the overlap index at any of the parameters $\mu$
listed up above. 
\footnote{On the other hand, if we decrease $\beta_{\rm W}$ for 
instance to $5.85$, the charge $Q_{\rm top}$ depends significantly
on the method of its determination \cite{BS}.}
Hence the results in the Tables are relevant
for the overlap index as well.

\begin{figure} 
\vspace*{-1cm} 
\begin{center}
 \includegraphics[angle=0,width=.49\linewidth]{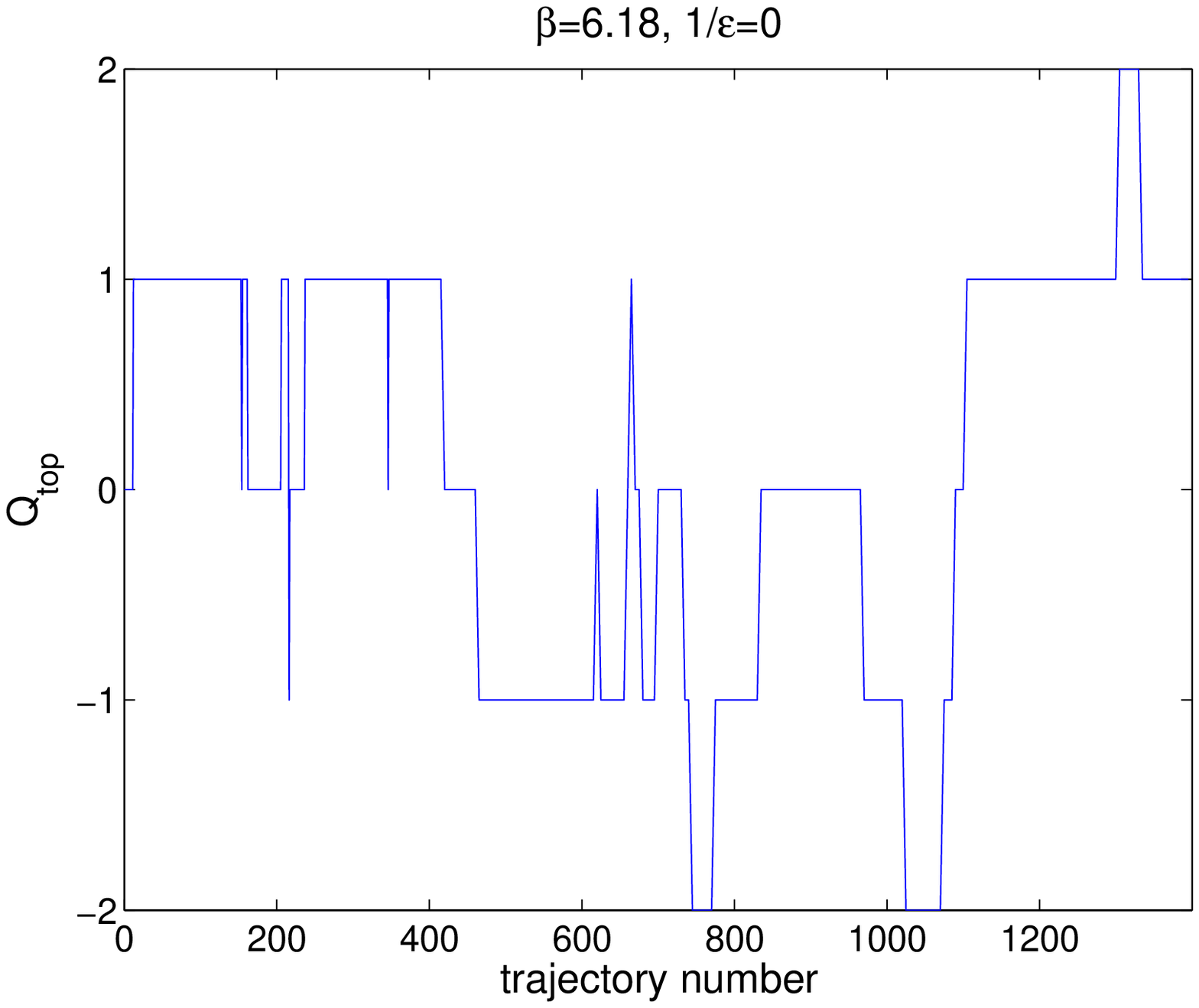}
 \includegraphics[angle=0,width=.49\linewidth]{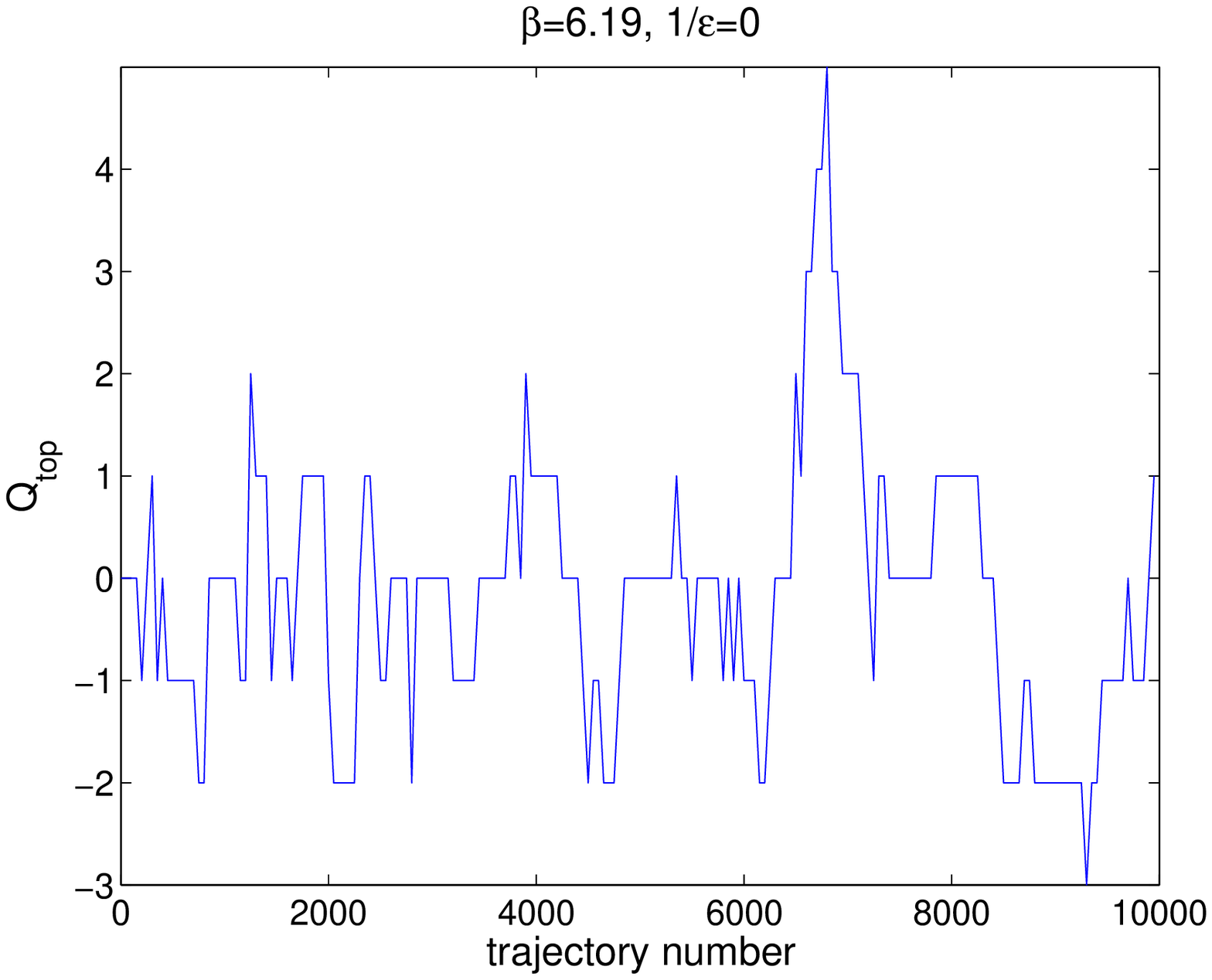}
 \includegraphics[angle=0,width=.49\linewidth]{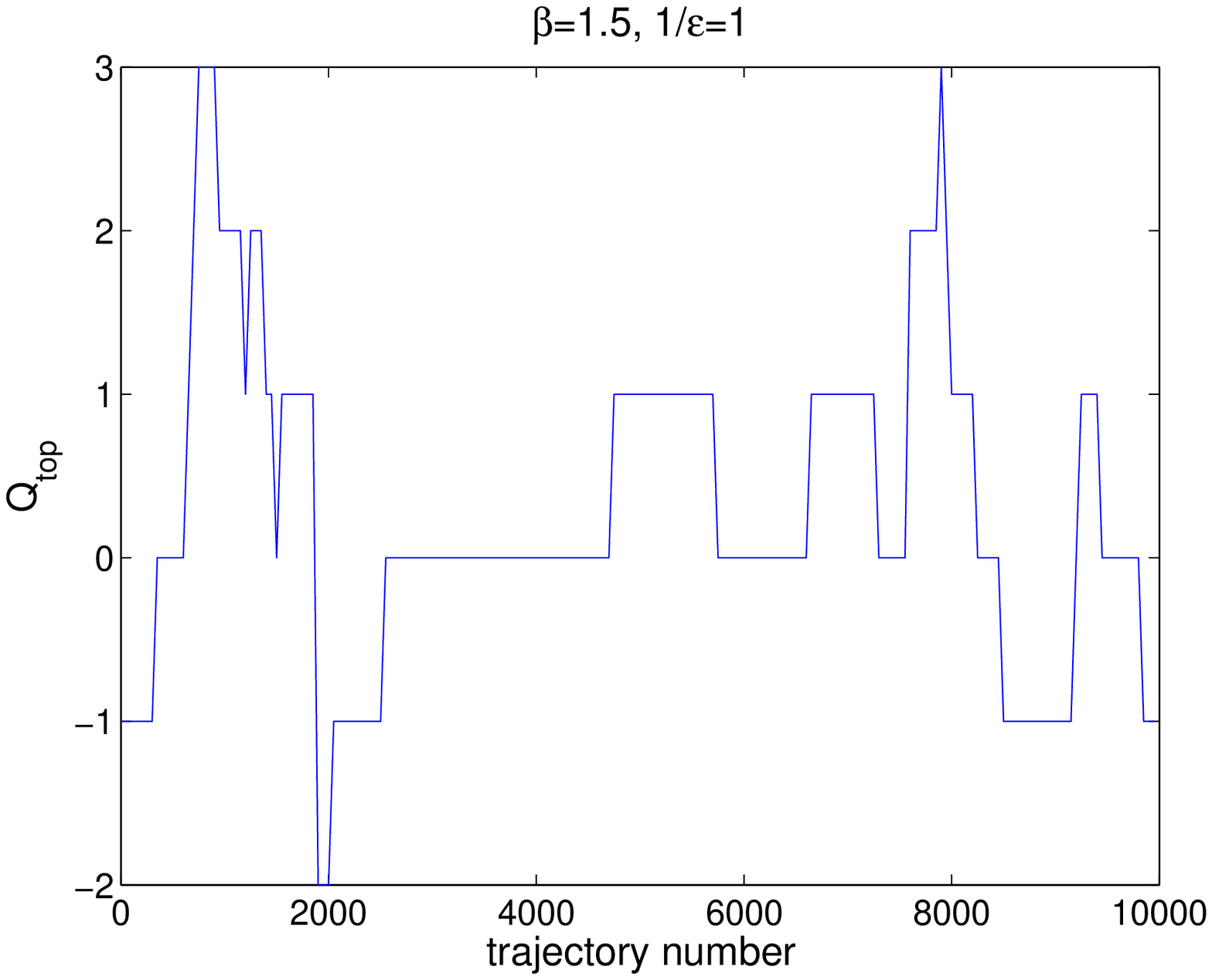}
 \includegraphics[angle=0,width=.49\linewidth]{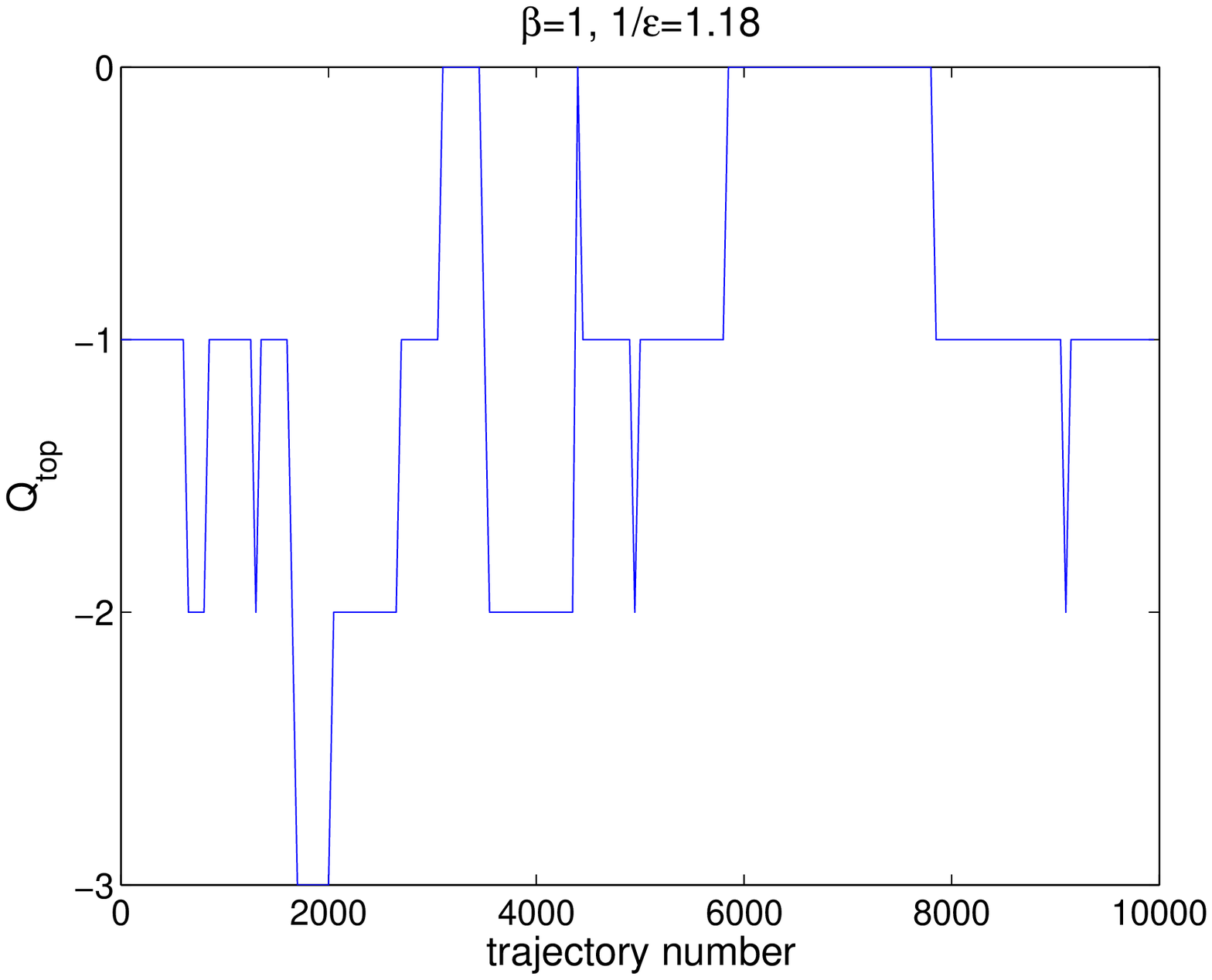}
 \includegraphics[angle=0,width=.49\linewidth]{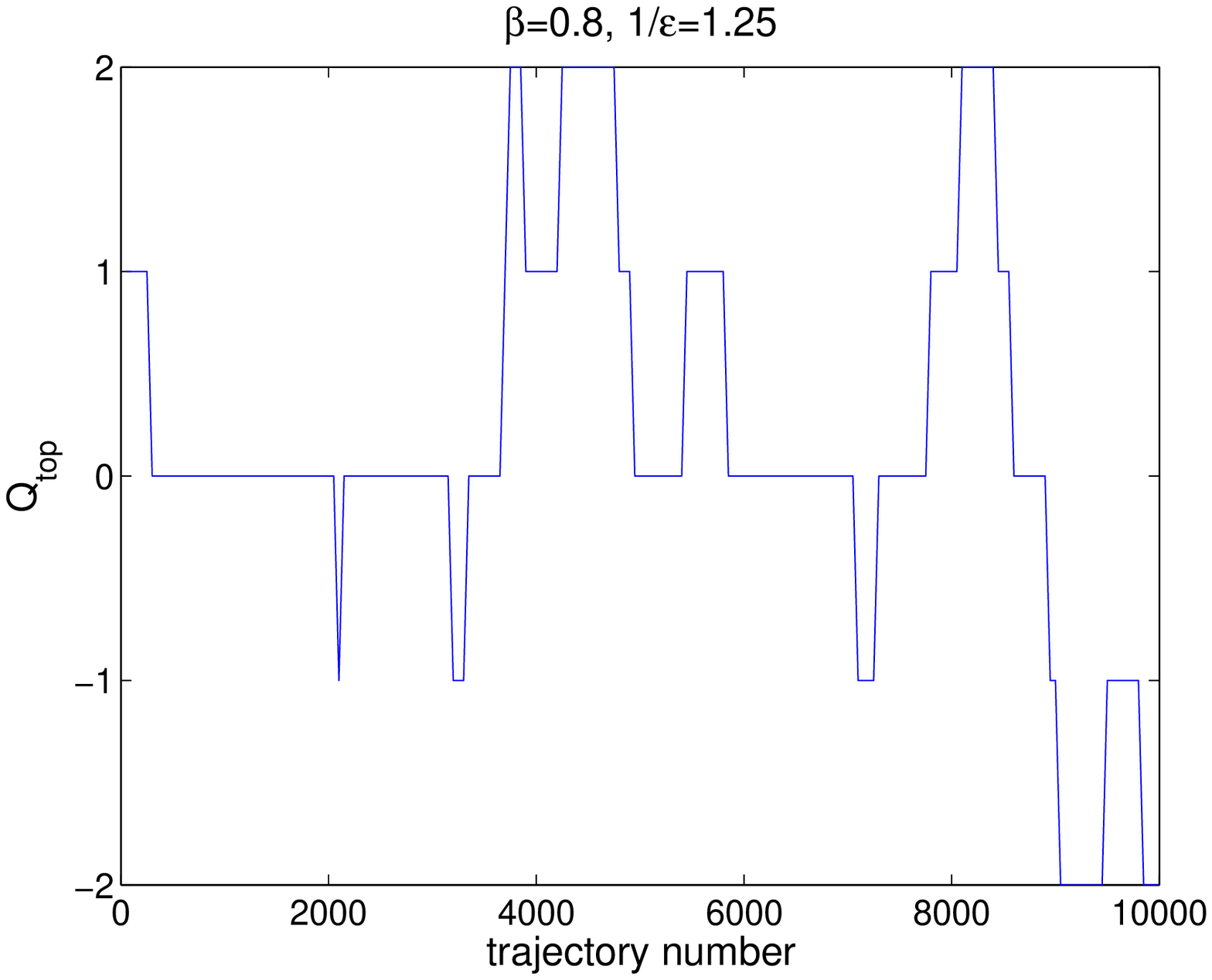}
 \includegraphics[angle=0,width=.49\linewidth]{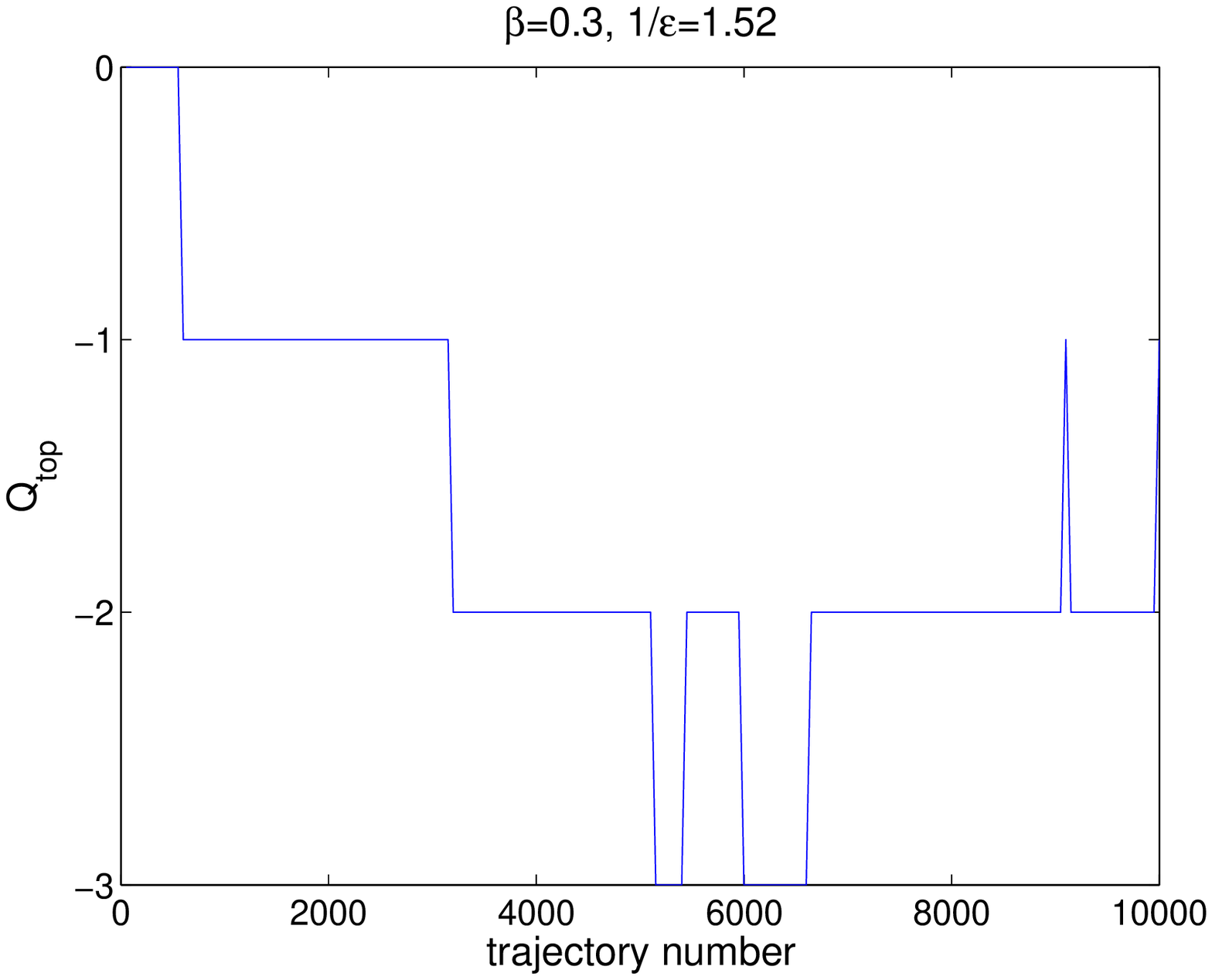}
\end{center}
\vspace{-5mm}
\caption{{\it Typical histories of the topological charge
for the actions $S_{\rm W}$ (on top) and
$S_{\varepsilon ,1}^{\rm hyp}$ at various combinations of
$\varepsilon$ and $\beta$, as in Table \ref{tab_hyp}. 
We show results obtained on a $16^{4}$ lattice with the
HMC step size $d \tau = 0.1$. We see that an increased 
$1/ \varepsilon$ keeps the charge more and more stable. \newline
The charge was measured by cooling
once in 50 trajectories, except for 
the plot at $\beta = 6.18$, where measurements were made every single or 
every 5 trajectories. Notice, on the other hand, that in the plot at 
$\beta = 6.19$ the assumption that the separation of measurements is much 
larger than the typical distance of topology changes is not justified, 
and the frequency of topological transitions based on this plot 
would be underestimated.}}
\label{tophisto}
\end{figure}

As a more direct illustration, we show typical histories of
the topological charge (defined by cooling) for different 
parameters in Figure \ref{tophisto}.
To measure the stability of the topological sector, we monitored the number
of charge changes normalised by the number of trajectories. 
We denote this parameter as the ``frequency of topological transitions'', 
$f_{\rm top}$, and give results in Tables \ref{tab_hyp}
and \ref{tab_exp}. 

As we already mentioned in Subsection 4.4, for a fixed
$d \tau$ the acceptance rate drops when we increase
$1/ \varepsilon$ up to $1.64$. Although this problem can be alleviated 
by reducing $d \tau$, it indicates that at this point the system
tends to run too often into forbidden regions, beyond the admissibility
cutoff. Therefore we also explored an action, which does not have
any strictly forbidden plaquette values. For instance, for the actions
in eqs.\ (\ref{powact}) and (\ref{expact}) the suppression of small
plaquette values rises smoothly. Our results for the exponential action
$S^{\rm exp}_{\varepsilon , 8}$
are collected in Table \ref{tab_exp}. We see that these actions do
allow us to render the topology somewhat more stable, and again small
HMC steps help us to keep the acceptance rate high.
However, it is difficult to fulfil the requirement $r_{0}/a \lsim 7$,
although we already chose extremely low values for $\beta$.
Therefore we did not push further into that direction.

Let us add some technical aspects about the evaluation of $f_{\rm top}$.
Although measuring the charges by cooling is rather cheap,
it could still not be evaluated on each trajectory (since we were
dealing with quite long histories). A reliable determination of $f_{\rm top}$
can only be done if the number of trajectories, which are skipped between
two measurements of the charge $Q_{\rm top}$, 
is much less than the typical number of
trajectories over which $Q_{\rm top}$ remains constant.
It turned out that for $S_{\rm W}$ it was sufficient to cool one 
configuration in 5 trajectories. For the gauge actions at $1/\varepsilon >0$, 
one configuration out of 50 trajectories was typically enough.

The error on $f_{\rm top}$ was estimated only in a crude way.
This is done by counting the transitions in 5 sub-histories and taking the
standard deviation from these 5 samples. The idea is inspired by the 
jack-knife method, but the difference is that the sub-histories have to
consist of contiguous elements.\\

\vspace*{-1mm}
Of course, we also need to consider the effect of $1/\varepsilon >0$ on the
autocorrelation time with respect to non-topological quantities.
One could be worried that an improved topological stability comes along with
a longer autocorrelation for other observables as well.
Our consideration of the plaquette value indicates the opposite:
its autocorrelation time {\em decreases} significantly with increasing
$1/ \varepsilon$, see Tables \ref{tab_hyp} and \ref{tab_exp}.

\begin{figure} 
%\vspace*{-1cm} 
\begin{center}
 \includegraphics[angle=0,width=.7\linewidth]{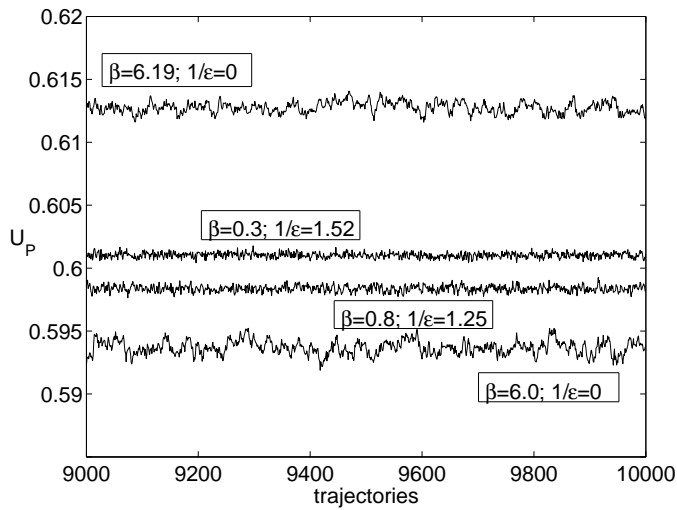}
\end{center}
\vspace{-5mm}
\caption{{\it Comparison of short portions of plaquette histories for 
different combinations of $\beta$ and $\varepsilon$.}}
\label{plaqfig}
\end{figure}

Another conceivable problem could be bad ergocity properties
even within one topological sector as  $1 / \varepsilon$ is switched on.
We checked this by performing simulations from independent starting points
and found that the mean plaquette values agree to a very high precision,
see Table \ref{plaqtab} and Figure \ref{plaqfig}.

\begin{table}
\begin{center}
\begin{tabular}{|l|c|l|c|c|}
\hline
$ ~~ \beta $ &  $1 /\varepsilon$ & $ ~~~~~ \langle U_{P} \rangle$ &    
$\tau^{\rm plaq}$ & starting point \\
\hline
\hline
6    &  0    &  0.59371(3)  &  9.2  &  cold  \\
\hline
6.19 &  0    &  0.61181(2)  &  7.2  &  cold  \\
\hline
0.8  & 1.25  &  0.598371(4) &  1.1  &  cold  \\
\hline
0.8  & 1.25  &  0.598372(4) &  1.1  &  $Q_{\rm top} = 1$ \\
\hline
0.8  & 1.25  &  0.598367(4) &  1.1  &  $Q_{\rm top} = 2$ \\
\hline
0.8  & 1.25  &  0.598369(4) &  1.0  &  $Q_{\rm top} = 3$ \\
\hline
0.3  & 1.52  &  0.601034(3) &  0.8  &  cold  \\
\hline
0.3  & 1.52  &  0.601028(3) &  0.8  &  $Q_{\rm top} = 1$ \\
\hline
\end{tabular}
\end{center}
\caption{\it Comparison of mean plaquette values $\langle U_{P} \rangle$
for different parameters and different starting points (out of 10 000 
trajectories in a volume $16^4$). The decreased plaquette autocorrelations 
lead to a much more precise determination of $\langle U_{P} \rangle$.
We see a remarkable agreement up to this high precision
for different starting points, even in different topological sectors.}
\label{plaqtab}
\end{table}

Of course, such tests should be extended to further observables, but our 
results for the plaquette value are encouraging.

\vspace*{-2mm}
\section{Conclusions}

The conservation of the topological charge can be implemented
in the lattice gauge action to some extent. There are various obstacles
preventing a strict implementation, such as scaling artifacts and the 
acceptance rate. Still, we succeeded in obtaining long 
sequences of a stable topological charge,
%(although it cannot be fixed for ever, of course). 
although it cannot be fixed strictly, as expected.
Our comparison to the behaviour of the 
plaquette value suggests that the topological autocorrelation time 
exceeds by far the autocorrelation of other observables.
This property facilitates the collection
of many configurations in a specific topological sector, for actions
which are perfectly acceptable.
We also observed that such actions do not seem to suffer from any
conceptual problems. In particular the resulting static potential
is fully consistent with the right continuum limit, and
it has relatively mild lattice artifacts.
We also showed that at least
a positive squared transfer matrix can be defined.

Moreover we found benefits of this action for the kernel
condition number of the overlap operator, which allows for a
somewhat faster evaluation with a fixed accuracy.
We finally remark that our findings are in accordance 
with recent results of Refs.\ \cite{JapQCD2}.
Further virtues, in particular in view of the simulations
with dynamical quarks, still remain to be explored.\\

\noindent
{\bf Acknowledgement} \ \ \ We would like to thank M.\ Creutz, H.\ Fukaya,
S.\ Hashimoto, M.\ Ilgenfritz, M.\ L\"{u}scher, M.\ M\"{u}ller-Preu\ss ker,
K.\ Ogawa, T.\ Onogi, R.\ Sommer and C. Urbach
for interesting discussions and helpful remarks.
This work was supported by the ``Deutsche Forschungsgemeinschaft'' 
through SFB/TR9-03.
S.\ Necco is supported by TMR, EC-Contract No.\ HPRNCT-2002-00311 (EURIDICE).
Part of the computations were performed
on the IBM p690 clusters of the ``Norddeutscher Verbund f\"ur Hoch- und 
H\"ochstleist\-ungsrechnen'' (HLRN) and at the Forschungszentrum J\"{u}lich.


\vspace*{-3mm}
\begin{thebibliography}{50}
\vspace*{-1mm}

\bibitem{Wilson} K.G.\ Wilson, {\it Phys.\ Rev.} {\bf D10} (1974) 2445.

\bibitem{MoMu} I.\ Montavay and G.\ M\"{u}nster,
``Quantum Fields on the Lattice'', Cambridge University Press
(Cambridge UK, 1994).

\bibitem{PW} P.\ Weisz, {\it Nucl.\ Phys.} {\bf B212} (1983) 1.

\bibitem{LW} M.\ L\"{u}scher and P.\ Weisz,
{\it Phys.\ Lett.} {\bf B158} (1985) 250;
{\it Commun.\ Math.\ Phys.} {\bf 97} (1985) 59.

\bibitem{Iwa} Y.\ Iwasaki, UTHEP-118.
Y.\ Iwasaki and T.\ Yoshi\'{e}, {\it Phys.\ Lett.} {\bf B143} (1984) 449.

\bibitem{DBW2} P.\ DeForcrand {\it et al.} (QCDTARO Collaboration),
{\it Nucl.\ Phys. (Proc.\ Suppl.)} {\bf B53} (1997) 938.
T.\ Takaishi, {\it Phys.\ Rev.} {\bf D54} (1996) 1050.

\bibitem{FPA} F.\ Niedermayer, P.\ R\"{u}fenacht and U.\ Wenger,
{\it Nucl. Phys.} {\bf B597} (2001) 413.

\bibitem{MLdyn} M.\ L\"{u}scher, {\tt hep-lat/0409106},
{\tt hep-lat/0509152}.
C.\ Urbach, K.\ Jansen, A.\ Shindler and U.\ Wenger, {\tt hep-lat/0506011}.

\bibitem{XPT} S.\ Weinberg, {\it Physica} {\bf A96} (1979) 327.
J.\ Gasser and H.\ Leutwyler, {\it Ann.\ Phys.\ (N.Y.)} {\bf 158} (1984) 142.

\bibitem{GW} P.H.\ Ginsparg and K.G.\ Wilson, 
{\it Phys. Rev.} {\bf D25} (1982) 2649.

\bibitem{Has} P.\ Hasenfratz, V.\ Laliena and F.\ Niedermayer, 
{\it Phys.\ Lett.} {\bf B427} (1998) 125.
P.\ Hasenfratz, {\it Nucl.\ Phys.} {\bf B525} (1998) 401. 

\bibitem{ML} M.\ L\"uscher, {\it Phys.\ Lett.} {\bf B428} (1998) 342.

\bibitem{Neu1} H.\ Neuberger, {\it Phys.\ Lett.} {\bf B417} (1998) 141;
{\it Phys.\ Lett.} {\bf B427} (1998) 353.

\bibitem{p-reg} J.\ Gasser and H.\ Leutwyler, 
{\it Phys. Lett.} {\bf B184} (1987) 83.

\bibitem{eps-reg} J.\ Gasser and H.\ Leutwyler,
{\it Phys.\ Lett.} {\bf B188} (1987) 477.
H.\ Neuberger, {\it Phys.\ Rev.\ Lett.} 
{\bf 60} (1988) 889;
{\it Nucl.\ Phys.} {\bf B300} (1988) 180.
P.\ Hasenfratz and H.\ Leutwyler, {\it Nucl.\ Phys.} 
{\bf B343} (1990) 241.
F.C.\ Hansen, {\it Nucl.\ Phys.} {\bf B345} (1990) 685.
F.C.\ Hansen and H.\ Leutwyler,
{\it Nucl.\ Phys.} {\bf B350} (1991) 201.
W.\ Bietenholz, {\it Helv.\ Phys.\ Acta} {\bf 66} (1993) 633.

\bibitem{LeuSmi} H.\ Leutwyler and A.\ Smilga, 
{\it Phys.\ Rev.} {\bf D46} (1992) 5607.
P.H.\ Damgaard {\it et al.}, {\it Nucl.\ Phys.} {\bf B629} (2002) 445,
{\it Nucl.\ Phys.} {\bf B656} (2003) 226.

\bibitem{sasa}  S.\ Prelovsek and K.\ Orginos,
{\it Nucl.\ Phys.\ (Proc.\ Suppl.)} {\bf B119} (2003) 822.

\bibitem{spec} W.\ Bietenholz, K.\ Jansen and S.\ Shcheredin, {\it JHEP}
{\bf 07} (2003) 033.
L.\ Giusti, M.\ L\"uscher, P.\ Weisz and H.\ Wittig,
{\it JHEP} {\bf 11} (2003) 023.
D.\ Galletly {\it et al.}, 
{\it Nucl.\ Phys.\ (Proc.\ Suppl.)} {\bf B129/130} (2004) 456.

\bibitem{toposus}  L.\ Giusti, M.\ L\"uscher, P.\ Weisz and H.\ Wittig,
{\it JHEP} {\bf 0311} (2003) 023.
L.\ Del Debbio and C.\ Pica, {\it JHEP} {\bf 0402} (2004) 003.
L.\ Del Debbio, L.\ Giusti and C.\ Pica, 
{\it Phys.\ Rev.\ Lett.} {\bf 94} (2005) 032003.

\bibitem{AA} W.\ Bietenholz, T.\ Chiarappa, K.\ Jansen, K.-I.\ Nagai,
and S.\ Shcheredin, {\it JHEP} {\bf 02} (2004) 023.

\bibitem{zeromode} L.\ Giusti, P.\ Hern\'{a}ndez, M.\ Laine, 
P.\ Weisz and H.\ Wittig, {\it JHEP} {\bf 01} (2004) 003.

\bibitem{LMA} L.\ Giusti, P.\ Hern\'{a}ndez, M.\ Laine, P.\ Weisz 
and H.\ Wittig, {\it JHEP} {\bf 04} (2004) 013.
L.\ Giusti and S.\ Necco,
{\it PoS(LAT2005)132} [{\tt hep-lat/0510011}].

\bibitem{BS} S.\ Shcheredin, Ph.D.\ thesis (Berlin, 2004) 
[{\tt hep-lat/0502001}].
W.\ Bietenholz and S.\ Shcheredin, 
{\it Rom.\ J.\ Phys.} {\bf 50} (2005) 249 [{\tt hep-lat/0502010}];
{\it PoS(LAT2005)138} [{\tt hep-lat/0508016}].
S.\ Shcheredin and W.\ Bietenholz,
{\it PoS(LAT2005)134} [{\tt hep-lat/0508034}].

\bibitem{Japeps} H.\ Fukaya, S.\ Hashimoto and K.\ Ogawa,
{\it Prog.\ Theor.\ Phys.} {\bf 114} (2005) 451;
{\it PoS(LAT2005)134} [{\tt hep-lat/0510049}].
K.\ Ogawa and S.\ Hashimoto,
{\it Prog.\ Theor.\ Phys.} {\bf 114} (2005) 609.

\bibitem{WenWit} J.\ Wennekers and H.\ Wittig,
{\it JHEP} {\bf 0509} (2005) 059.
L.\ Giusti, P.\ Hern\'{a}ndez, M.\ Laine, C.\ Pena, J.\ Wennekers and 
H.\ Wittig, {\it PoS(LAT2005)344} [{\tt hep-lat/0510033}].

\bibitem{DamNish} P.H.\ Damgaard and S.M.\ Nishigaki,
{\it Nucl.\ Phys.} {\bf B518} (1998) 495;
{\it Phys.\ Rev.} {\bf D63} (2001) 045012.

\bibitem{nonplaq}  K.\ Orginos,
{\it Nucl.\ Phys.\ (Proc.\ Suppl.)} {\bf 106} (2002) 721.
Y.\ Aoki {\it et al.}, 
%T. Blum, N. Christ, C. Cristian, C. Dawson, T. Izubuchi, 
%G. Liu, R. Mawhinney, S. Ohta, K. Orginos, A. Soni, L. Wu
{\it Phys.\ Rev.} {\bf D69} (2004) 074504.

\bibitem{HJL} P.\ Hern\'{a}ndez, K.\ Jansen and M.\ L\"{u}scher,
{\it Nucl.\ Phys.} {\bf B552} (1999) 363.

\bibitem{Neu2} H.\ Neuberger, {\it Phys.\ Rev.} {\bf D61} (2000) 085015.

\bibitem{Nagao} K.\ Nagao, {\tt hep-th/0509034}.

\bibitem{luschact}  M.\ L\"{u}scher, {\it Nucl.\ Phys.} {\bf B549} (1999) 
295; {\it Nucl.\ Phys.} B568 (2000) 162.

\bibitem{FuOn} H.\ Fukaya and T.\ Onogi, Phys.\ Rev.\ {\bf D68} (2003) 074503;
{\it Phys.\ Rev.} {\bf D70} (2004) 054508.

\bibitem{Creutz} M.\ Creutz, {\it Phys.\ Rev.} {\bf D70} (2004) 091501.

\bibitem{prelim} S.\ Shcheredin, W.\ Bietenholz, K.\ Jansen, K.-I.\ Nagai, 
S.\ Necco and L. Scorzato, 
{\it Nucl.\ Phys.\ (Proc.\ Suppl.)} {\bf B140} (2005) 779 
[{\tt hep-lat/0409073}].
W. Bietenholz, K. Jansen, K.-I. Nagai, S. Necco,
L. Scorzato and S. Shcheredin, {\it AIP Conf.\ Proc.} {\bf 756} (2005) 248
[{\tt hep-lat/0412017}].
K.-I.\ Nagai, K.\ Jansen, W.\ Bietenholz, L.\ Scorzato, 
S.\ Necco and S.\ Shcheredin, {\it PoS(LAT2005)283} [{\tt hep-lat/0509170}].

\bibitem{JapQCD} H.\ Fukaya, T.\ Onogi, S.\ Hashimoto, T.\ Hirohashi
and K.\ Ogawa, {\it PoS(LAT2005)134} [{\tt hep-lat/0509184}].

\bibitem{JapQCD2} H.\ Fukaya, S.\ Hashimoto, T.\ Hirohashi, 
H.\ Matsufuru, K.\ Ogawa, 
and T.\ Onogi, {\it PoS(LAT2005)123} [{\tt hep-lat/0510095}].
H.\ Fukaya, S.\ Hashimoto, T.\ Hirohashi, K.\ Ogawa, 
and T.\ Onogi, {\tt hep-lat/0510116}.

\bibitem{lHMC} P.\ Marenzoni, L.\ Pugnetti and P.\ Rossi, 
{\it Phys.\ Lett.} {\bf B315} (1993) 152.

\bibitem{APE} M.\ Albanese {\it et al.}, {\it Phys.\ Lett.} {\bf 192B} 
(1987) 163.

\bibitem{GSW} M.\ Guagnelli, R.\ Sommer and H.\ Wittig, 
{\it Nucl.\ Phys.} {\bf B535} (1998) 389.

\bibitem{NeSo} S.\ Necco and R.\ Sommer, {\it Nucl.\ Phys.} {\bf B622} 
(2002) 328.

\bibitem{RSo} R.\ Sommer, {\it Nucl.\ Phys.} {\bf B411} (1994) 839.

\bibitem{SNe} S.\ Necco, {\it Nucl.\ Phys.} {\bf B683} (2004) 137;
Ph.D.\ thesis (Berlin, 2004) [{\tt hep-lat/0306005}].

\bibitem{NoManton} H.\ Grosse and H.\ Kuhnelt, {\it Nucl. Phys.} {\bf B205}
(1982) 273.

\bibitem{Manton} N.\ Manton, {\it Phys.\ Lett.} {\bf B96} (1980) 328.

\bibitem{ovimp} W.\ Bietenholz,
{\it Eur.\ Phys.\ J.} {\bf C6} (1999) 537;
{\it Nucl.\ Phys.} {\bf B644} (2002) 223.
W. Bietenholz and I.\ Hip, {\it Nucl.\ Phys.} {\bf B570}
(2000) 423 [{\tt hep-lat/9902019}].
T.\ DeGrand, {\it Phys.\ Rev.} {\bf D63} (2001) 034503.
W.\ Kamleh, D.H.\ Adams, D.B.\ Leinweber and A.G. Williams,
{\it Phys.\ Rev.} {\bf D66} (2002) 014501.
S.\ D\"{u}rr, C.\ Hoelbling and U.\ Wenger, {\tt hep-lat/0506027}.

\bibitem{cool} E.-M.\ Ilgenfritz, M.L.\ Laursen, G.\ Schierholz,
M.\ M\"{u}ller-Preu\ss ker and H.\ Schiller, 
{\it Nucl.\ Phys.} {\bf B268} (1986) 693.

\end{thebibliography}
\end{document}